\documentclass[a4paper,11pt]{article}

\usepackage{amsmath,amsthm}

\usepackage[T1]{tipa}

\usepackage{wasysym}
\usepackage{amsfonts,cite}

\usepackage{bm} 

\usepackage[greek,english]{babel}

\newcommand{\be}{\begin{equation}}
\newcommand{\ee}{\end{equation}}
\newcommand{\bea}{\begin{eqnarray}}
\newcommand{\eea}{\end{eqnarray}}

\newcommand{\lb}{\mbox{\boldmath{$\ell$}}}
\newcommand{\kb}{\mbox{\boldmath{$k$}}}
\newcommand{\nb}{\mathbf{n}}
\newcommand{\mb}[1]{{{\mathbf m}_{(#1)}}}

\newcommand{\mdigamma} {\text{\foreignlanguage{greek}{\ddigamma}}}
\newcommand{\yogh} {\text{\textit{\textyogh}}}

\newcommand{\M}[3] {{\stackrel{#1}{M}}_{{#2}{#3}}}

\newcommand{\eb}{\mathbf {e}}

\newcommand{\diag}{{\mathrm{diag}}}
\newcommand{\pul}{{\textstyle{\frac{1}{2}}}}
\def \bl {\mbox{\boldmath{$\ell$}}}
\def \bn {\mbox{\boldmath{$n$}}}

\def \l {\ell}

\newcommand{\del}{{\delta}}
\newcommand{\Om}{\Omega} 
\newcommand{\Ps}{\Psi}

\newtheorem{theorem}{Theorem}

\newtheorem{prop}{Proposition}

\newcommand{\Phis}{\Phi^\mathrm{S}} 
\newcommand{\Phia}{\Phi^\mathrm{A}} 

\newcommand{\tho}{{\textrm{\thorn}}}
\newcommand{\rhob}{\bm{\rho}}

\newcommand{\Phib}{{\bm{\Phi}}}

\newcommand{\Sb}{{\mathbf{S}}}
\newcommand{\Id}{{\mathbf{1}}}

\newcommand{\beqn}{\begin{eqnarray}}
\newcommand{\eeqn}{\end{eqnarray}}
\newcommand{\pa}{\partial}
\newcommand{\ba}{\begin{array}}
\newcommand{\ea}{\end{array}}

\newcommand{\beqnn}{\begin{eqnarray*}}
\newcommand{\eeqnn}{\end{eqnarray*}}

\def \WDS #1 {\mbox{$\Phi_{#1}^{S}$}}
\def \WDA #1 {\mbox{$\Phi_{#1}^{A}$}}
\def \WD #1 {\mbox{$\Phi_{#1}$}}
\def \kd #1 {\delta_{#1}}

\def \BEAH {\begin{eqnarray*}}
\def \EEAH {\end{eqnarray*}}
\def \BEA {\begin{eqnarray}}
\def \EEA {\end{eqnarray}}
\def \BDM {\begin{displaymath}}
\def \EDM {\end{displaymath}}

\def \pul {{{\footnotesize{\frac{1}{2}}}}}

\numberwithin{equation}{section}

\title{On a five-dimensional version of the Goldberg-Sachs theorem}
\author{Marcello Ortaggio$^a$, Vojt\v ech Pravda$^a$, Alena Pravdov\'a$^a$ and Harvey S. Reall$^b$\\[3mm]
        {\small\it ${}^a$ Institute of Mathematics, Academy of Sciences, \v Zitn\'a 25, 115 67 Prague 1, Czech Republic}\\
        {\small\it ${}^b$ DAMTP, Centre for Mathematical Sciences, University of Cambridge,}\\
        {\small\it Wilberforce Road, Cambridge, CB3 0WA, United Kingdom}\\[1mm]
        {\small\tt ortaggio@math.cas.cz, pravda@math.cas.cz,}\\
        {\small\tt pravdova@math.cas.cz, hsr1000@cam.ac.uk}}
\date{}
\begin{document} 
\maketitle

\begin{abstract}
 Previous work has found a higher-dimensional generalization of the ``geodesic part'' of the Goldberg-Sachs theorem. We investigate the generalization of the ``shear-free part'' of the theorem. A spacetime is defined to be algebraically special if it admits a multiple Weyl Aligned Null Direction (WAND). The algebraically special property restricts the form of the ``optical matrix'' that defines the expansion, rotation and shear of the multiple WAND. {After working out some general constraints that hold in arbitrary dimensions,} we determine  necessary algebraic conditions on the optical matrix of a multiple WAND in a five-dimensional Einstein spacetime. We prove that one can choose an orthonormal basis to bring the $3\times 3$ optical matrix to one of three  canonical forms, each involving two parameters, and we discuss the existence of an ``optical structure'' within these classes. Examples of solutions corresponding to each form are given. We give an example which demonstrates that our necessary algebraic conditions are not sufficient for a null vector field to be a multiple WAND, in contrast with the 4d result.
\end{abstract}

\section{Introduction}

The Goldberg-Sachs (GS) theorem \cite{GolSac62,Stephanibook} states: {\it in an Einstein spacetime\footnote{
An Einstein spacetime is a solution of the vacuum Einstein equation allowing for a cosmological constant, { i.e., in $d$ dimensions, $R_{ab}=(R/d)g_{ab}$}.} which is not conformally flat, a null vector field is a repeated principal null direction (of the Weyl tensor) if, and only if, it is geodesic and shear-free}. This theorem plays an important role in the study of solutions of the Einstein equation with an algebraically special Weyl tensor. In particular, it is the first step in exploiting the algebraically special property to solve the Einstein equation, which is how the Kerr solution was discovered \cite{Kerr63}.

Given the interest in higher-dimensional solutions of the Einstein equation, it is natural to attempt to use the algebraically special property to obtain solutions. Therefore it seems desirable to obtain a higher-dimensional generalization of the Goldberg-Sachs theorem. An algebraic classification of the Weyl tensor in higher dimensions was given in Ref. \cite{Coleyetal04}. It is based on the notion of Weyl Aligned Null Directions (WANDs). In 4d, a WAND is the same as a principal null direction. The generalization of a repeated principal null direction is called a {\it multiple} WAND. We define a spacetime to be algebraically special if it admits a multiple WAND.

Simple examples reveal that the GS theorem does not extend in an obvious way to higher dimensions. Consider the Einstein spacetime $dS_3 \times S^2$ (where $dS_3$ denotes 3d de Sitter spacetime). For this spacetime, {\it any} null vector field tangent to $dS_3$ is a multiple WAND \cite{GodRea09}. (This demonstrates that (multiple) WANDs need not be discrete in higher dimensions.) But not all such vector fields are geodesic. Hence the ``geodesic part'' of the GS theorem does not generalize immediately to higher dimensions {\cite{PraPraOrt07}}. However, this example also admits multiple WANDs which {\it are} geodesic. It turns out that this behaviour is generic. Ref. \cite{DurRea09} proved that {\it an Einstein spacetime admits a multiple WAND if, and only if, it admits a geodesic multiple WAND}. Hence there is no loss of generality in restricting attention to geodesic multiple WANDs.

The ``shearfree'' part of the GS theorem also does not generalize immediately to higher dimensions {\cite{MyePer86,FroSto03,Pravdaetal04,OrtPraPra07,PraPraOrt07}}. Consider a black brane solution given by the product of the 4d Schwarzschild solution with some flat directions. The repeated principal null direction of the Schwarzschild solution is a multiple WAND in this spacetime. This multiple WAND is geodesic, but shearing because it expands in the Schwarzschild directions and not in the flat directions. Hence a higher-dimensional generalization of the GS theorem must allow for non-vanishing shear. 

To explain what we mean by a ``higher-dimensional generalization of the GS theorem'', consider a null geodesic vector field $\lb$ and introduce a set of $d-2$ orthonormal spacelike vectors $\mb{i}$ that are orthogonal to $\lb$. The expansion, rotation and shear of $\lb$ are defined as the trace, antisymmetric part, and traceless symmetric part, of the $(d-2) \times (d-2)$ {\it optical matrix}
\be
 \rho_{ij} = m_{(i)}^\mu m_{(j)}^\nu \nabla_\nu \ell_\mu . \label{expmatrix}
\ee
The most desirable form of a higher-dimensional generalization of the GS theorem, would be a statement of necessary and sufficient {\it algebraic} conditions on $\rho_{ij}$ for a geodesic null vector field $\lb$ to be a multiple WAND. However, it is not clear whether or not such a theorem exists in higher dimensions. There are certainly necessary conditions on $\rho_{ij}$ that follow from the multiple WAND condition. There are also conditions on $\rho_{ij}$ that are sufficient for $\lb$ to be a multiple WAND. But maybe there are no conditions which are both necessary and sufficient. In fact, this is what we will show below. But first we will summarize the known sufficient or necessary conditions on $\rho_{ij}$ for $\lb$ to be a multiple WAND.

{\it Sufficient} conditions on $\rho_{ij}$ for $\lb$ to be a multiple WAND in an Einstein spacetime have been obtained for Kundt spacetimes, defined by $\rho_{ij}=0$ (i.e. vanishing expansion, rotation and shear), and Robinson-Trautman spacetimes, defined by $\rho_{ij} = (d-2)^{-1} \rho_{kk} \delta_{ij}$ (i.e. vanishing rotation and shear, non-vanishing expansion). For both of these cases, $\lb$ must be a multiple WAND \cite{OrtPraPra07,PodOrt06}.

{\it Necessary} conditions on $\rho_{ij}$ following from the multiple WAND condition have been obtained as follows. 
Ricci flat  spacetimes of type N were considered in Ref. \cite{Pravdaetal04}.  
It was shown that the multiple WAND defined by the type N property must be geodesic, and the optical matrix must have rank 2. A basis can be found for which this matrix is zero everywhere except in the leading $2\times 2$ block, which has the form
\be
\label{2dblock}
 b \left( \begin{array}{cc} 1 & a \\ -a & 1 \end{array} \right).
\ee
This is equivalent to the shearfree condition if $d=4$ but not if $d>4$. Ref. \cite{Pravdaetal04} also found that the same form applies to type III vacuum spacetimes that either (i) are five-dimensional; or (ii) satisfy a certain genericity condition; or (iii) have a {\it non-twisting} multiple WAND, i.e., one with vanishing rotation (in which case $a=0$). As pointed out e.g. in \cite{Durkeeetal10}, generalization of the above type N and III results from the Ricci-flat case to the Einstein case is straightforward.

Results have also been obtained for the case of a Kerr-Schild spacetime. Such a spacetime has a metric of the form
\be
 g_{\mu \nu} = \bar{g}_{\mu \nu} + H k_\mu k_\nu,
\ee
where $\bar{g}_{\mu \nu}$ is a metric of constant curvature, $\kb$ is null with respect to $\bar{g}_{\mu \nu}$ (which implies that it is null also with respect to $g_{\mu \nu}$), and $H$ is a function. This class of spacetimes includes Myers-Perry black holes. Einstein Kerr-Schild spacetimes are algebraically special for any $d \ge 4$, with $\kb$ being a geodesic multiple WAND \cite{OrtPraPra09,MalPra11GKS}. For these spacetimes, it has been shown that one can choose the basis vectors $\mb{i}$ so that $\rho_{ij}$ is block diagonal, with the diagonal consisting of a set of $2\times 2$ blocks of the form (\ref{2dblock}) (the parameters $a,b$ can vary from block to block, satisfying the optical constraint discussed below), followed by a set of identical $1 \times 1$ blocks and then zeros 
\cite{OrtPraPra09,MalPra11GKS}. Similar results holding for asymptotically flat type II spacetimes are mentioned in Section~\ref{Sec_OC}.

In this paper, we will investigate the {\em necessary} conditions on $\rho_{ij}$ for $\lb$ to be a multiple WAND in a five-dimensional Einstein spacetime of type II (by ``type II'' we include also type D, and refer instead to ``genuine type II'' in the few cases when the type is II and not D and the distinction is important). When our results are combined with the results for type III and type N spacetimes discussed above, one obtains results for general algebraically special spacetimes. The main result (incorporating the "geodesic" result of Ref. \cite{DurRea09}) can be summarized as:

\begin{theorem}
\label{th_GS}
 In a 5d algebraically special Einstein spacetime that is not conformally flat, there exists a geodesic multiple WAND $\lb$ and one can choose the orthonormal basis vectors $\mb{i}$ so that the optical matrix of $\lb$ takes one of the forms 
\bea
\label{5dform1}
 &&i) \ \ \ \  b\left( \begin{array}{ccc} 1 & a & 0 \\ -a & 1 & 0 \\ 0 & 0& 1+a^2 \end{array} \right), 
 \\
 \label{5dform2}
 &&ii) \ \ \ \ b\left( \begin{array}{ccc} 1 & a & 0 \\ -a & 1 & 0 \\ 0 & 0& 0 \end{array} \right),\qquad 
 \\
\label{5dform3}
 &&iii) \ \ \ \ b\left( \begin{array}{ccc} 1 & a & 0 \\ -a & -a^2 & 0 \\ 0 & 0& 0 \end{array} \right) 
 .
\eea
 If the spacetime is type III or type N then the form must be $ii)$.
\end{theorem}
The only matrix belonging to more than one of the above classes is $\rho_{ij}=0$.
This can be seen by looking at the eigenvalues of $\rho_{(ij)}$.  Of course $a,b$ are functions that vary in spacetime. 

A $3 \times 3$ orthogonal matrix has 3 parameters so one can eliminate at most 3 parameters from a $3 \times 3$ matrix by a change of orthonormal basis. Hence the canonical form of a general optical matrix will have $9-3=6$ parameters. Our theorem shows that the multiple WAND condition reduces this to the $2$ parameters $a,b$, {thus giving algebraic relations between the expansion, shear and twist of the multiple WAND}. (In 4d the corresponding result is a reduction from 3 to 2 parameters.) 

An example of an algebraically special solution for which $\rho_{ij}$ can be brought to the form (\ref{5dform1}) is the Myers-Perry  \cite{MyePer86} black hole solution {(cf.~\cite{PraPraOrt07})}. The Kerr black string (i.e. the product of the 4d Kerr solution with a flat direction) gives $\rho_{ij}$ which can be brought to the form (\ref{5dform2}). We will show {in Section~\ref{Sec_examples}} that a {specific} geodesic multiple WAND in $dS_3 \times S^2$ gives an example of the form~(\ref{5dform3}). These examples reveal that Theorem 1 is sharp: one cannot obtain further restrictions on the parameters $a,b$ (except perhaps in the form of inequalities). 

As mentioned above, our Theorem gives necessary conditions for $\lb$ to be a geodesic multiple WAND but these conditions are not sufficient. In Section~\ref{subsec_counterex} we give an example of an Einstein spacetime that is not algebraically special yet admits a geodesic null vector field with an optical matrix of the form (iii) above. 

We have used a definition of ``algebraically special'' based on the classification of the Weyl tensor of Ref. \cite{Coleyetal04}. Other definitions of ``algebraically special'' have been proposed in higher dimensions. For the special case of 5 dimensions, Ref. \cite{DeSmet02} performed a spinorial classification of the Weyl tensor. In this approach, null directions do not play a privileged role and so it seems unlikely that there is any generalization of the GS theorem. Ref. \cite{Taghavi-Chabert11} proposed a definition of algebraically special which, subject to a genericity assumption that the Weyl tensor is not ``too special'', implies the existence of an ``optical structure'' generalizing the notion of a 4d null geodesic congruence with vanishing shear. 
A spacetime that is algebraically special in the sense of Ref. \cite{Taghavi-Chabert11} is also algebraically special in our sense but the converse is not true. Hence the definition of Ref.~\cite{Taghavi-Chabert11} is more restrictive than the one used here. {The relation between our work and the results of \cite{Taghavi-Chabert11} will be discussed in Section~\ref{subsec_TC}.

This paper is organized as follows: Section \ref{prelim} introduces notation. In Section \ref{Sec_constraints} we discuss the algebraic constraints governing type II Einstein spacetimes. 
In Section \ref{sec_5D} we restrict to 5d and study the conditions on the optical matrix  that allow for a non-trivial solution of these constraints. By examining all such forms of the optical matrix we prove Theorem~\ref{th_GS}. 
{In Section~\ref{sec_integrab} we discuss consequences of our results in terms of the integrability properties of certain distributions and, in particular, the existence of an optical structure, and we provide a comparison with the results of \cite{Taghavi-Chabert11}. Section~\ref{Sec_examples} discusses some examples of algebraically special solutions corresponding to the three canonical forms of the optical matrix
and a counterexample to the converse of Theorem~\ref{th_GS}.  Section~\ref{sec_discussion} is devoted to discussion of the results.
Finally, in Appendix~\ref{app} 
we present a result concerning non-twisting multiple WANDs in an arbitrary number of dimensions.

\section{Preliminaries}

\label{prelim}

Throughout this paper we use higher-dimensional GHP formalism of Ref. \cite{Durkeeetal10}. For convenience let us   summarize necessary notation.  We employ a null frame
\begin{equation}
  \{\lb \equiv \eb_{(0)}=\eb^{(1)},
    \nb \equiv \eb_{(1)} = \eb^{(0)},
    \mb{i}\equiv\eb_{(i)} = \eb^{(i)} \}
\end{equation}
with indices $i,j,k,\ldots$ running from $2$ to $d-1$. The vector fields  $\lb$ and $\nb$ are null and the orthonormal spacelike vector fields $\mb{i}$  obey $\lb \cdot \nb =1$, $\mb{i} \cdot \mb{j} = \delta_{ij}$ and  $\lb \cdot \mb{i} = 0 = \nb \cdot \mb{i}.$

The GHP formalism is designed to maintain covariance with respect to two special types of Lorentz transformation. These are {\it boosts}, defined by 
\be
 \lb \rightarrow \lambda \lb, \qquad \nb \rightarrow \lambda^{-1} \nb, \qquad \mb{i} \rightarrow \mb{i} 
\ee
for some function $\lambda$, and {\it spins}, defined by a (position-dependent) rotation of the spatial basis vectors. A set of tensor components $T_{i_1 \ldots i_s}$ is said to be a {\it GHP tensor} of spin $s$ and boost weight $b$ if, under a spin, it transforms as a Cartesian tensor of rank $s$, and under a boost $T_{i_1 \ldots i_s} \rightarrow \lambda^b T_{i_1 \ldots i_s}$. 

The frame components of the Weyl tensor   with respect to the null basis are
\begin{equation}
  C_{ab...c} = e^\mu_{(a)} e^\nu_{(b)} ... e^\rho_{(c)} C_{\mu\nu...\rho},
\end{equation}
where $d$-dimensional coordinate indices $\mu,\nu,\ldots$ and frame indices $a,b,\ldots$ take values from $0$ to $d-1$. The notation of Ref. \cite{Durkeeetal10} for various frame components of the Weyl tensor is summarized in  Table \ref{tab:weyl}.
\begin{table}[ht]
  \begin{center}
  \begin{tabular}{|c|c|c|l|}
    \hline $b$ & Compt. & Notation & Identities  \\\hline
    $2$ & $C_{0i0j}$& $\Om_{ij}$   & $\Om_{ij} = \Om_{ji}$, $\Om_{ii}=0$ \\\hline
    $1$ & $C_{0ijk}$& $\Ps_{ijk}$  &  $\Ps_{ijk} = -\Ps_{ikj}$, $\Ps_{[ijk]}=0$ 
                                                            \\
      & $C_{010i}$& $\Ps_{i}$    &  $\Ps_i = \Ps_{kik}$.  \\\hline
    $0$ & $C_{ijkl}$& $\Phi_{ijkl}$&  $\Phi_{ijkl} = \Phi_{[ij][kl]} = \Phi_{klij}$, $\Phi_{i[jkl]}=0$
                                                          \\
      & $C_{0i1j}$& $\Phi_{ij}$  &  $\Phi_{(ij)} \equiv \Phis_{ij} = -\pul\Phi_{ikjk}$  \\
      & $C_{01ij}$& $2\Phia_{ij}$&  $\Phia_{ij} \equiv \Phi_{[ij]}$  \\
      & $C_{0101}$& $\Phi$       &  $\Phi=\Phi_{ii}$  \\\hline
    $-1$& $C_{1ijk}$& $\Ps'_{ijk}$ &  $\Ps'_{ijk} = -\Ps'_{ikj}$, $\Ps'_{[ijk]}=0$
                                                             \\
      & $C_{101i}$& $\Ps'_{i}$   &  $\Ps'_i = \Ps'_{kik}$.  \\\hline
    $-2$& $C_{1i1j}$& $\Om'_{ij}$  &  $\Om'_{ij} = \Om'_{ji}$, $\Om'_{ii}=0$ \\
\hline
  \end{tabular}
    \caption{Weyl tensor components sorted by boost weight $b$ in a $d\geq4$-dimensional spacetime. (c.f. Refs. \cite{Coleyetal04,Durkeeetal10})\label{tab:weyl}}
  \end{center}
\end{table}

The GHP formalism uses derivative operators which act covariantly on GHP tensors \cite{Durkeeetal10}. In this paper we will need only the following modification of the directional derivative along $\lb$:
\be
 \tho  T_{i_1 \ldots i_s} = \left( \ell^\mu \partial_\mu   -b n^\nu \ell^\mu \nabla_\mu \ell_\nu \right)  T_{i_1 \ldots i_s} + \sum_{r=1}^s \left( m_{(i_r)}^\nu \ell^\mu \nabla_\mu m_{(k)\nu} \right) T_{i_1 \ldots i_{r-1} k i_{r+1} \ldots i_s} .
 \ee
In {the following} we will assume that spacetime is algebraically special, i.e., it admits a multiple WAND. We choose $\lb$ to be aligned with the multiple WAND. Then the algebraically special condition is \cite{Coleyetal04}
\be
\Om_{ij} = \Ps_{ijk} = \Ps_{i} = 0.
\ee
{Thanks to the results of \cite{DurRea09}, with no loss of generality we can choose $\lb$ to be geodesic, i.e., $\kappa_i=0$.}
The following notation is used for the symmetric and antisymmetric parts of the optical matrix $\rho_{ij}$:  
\be
 S_{ij} = \rho_{(ij)}, \qquad A_{ij} = \rho_{[ij]}. 
\ee
We  sometimes drop spatial indices $i,j,\dots$ and use boldface  $\rhob $ instead of $\rho_{ij}$. In contrast,
the trace of the optical matrix $\rho_{ii}$ is denoted by $\rho$:
\be
 \rho = \rho_{ii} .
\ee

\section{Algebraic constraints for type II spacetimes in arbitrary dimension}
\label{Sec_constraints}

\subsection{Differentiating Bianchi components}

The components of the Bianchi identity are written out in Refs.  \cite{Pravdaetal04,Durkeeetal10}. For algebraically special spacetimes, some of these equations reduce to purely algebraic constraints. In particular, Eqs. (B4) and (B8) of \cite{Durkeeetal10}, in any number of dimensions, 
reduce to 
\bea
2\Phia_{[jk|}\rho_{i|l]} -2\Phi_{i[j}\rho_{kl]} + \Phi_{im [jk|}\rho_{m|l]}  = 0, \label{B4} \\
\Phi_{kj} \rho_{ij} - \Phi_{jk} \rho_{ij} + \Phi_{ij} \rho_{kj} - \Phi_{ji} \rho_{jk} 
       + 2 \Phi_{ij} \rho_{jk} - \Phi_{ik} \rho  + \Phi \rho_{ik} + \Phi_{ijkl} \rho_{jl} = 0 \label{B8},
\eea
respectively. Eq. \eqref{B8} is traceless and its
symmetric and antisymmetric parts  read
\bea
\left( 2 \Phi_{kj} - \Phi_{jk} \right) S_{ij} + \left( 2 \Phi_{ij} - \Phi_{ji} \right) S_{jk}  - \Phis_{ik} \rho + \Phi S_{ik} + \Phi_{ijkl} S_{jl} =0, \label{B8sym}\\
\Phi_{jk} A_{ji} + \Phi_{ji} A_{kj} + \Phi_{ij} \rho_{jk} -\Phi_{kj} \rho_{ji} +  \Phia_{k i} \rho + \Phi A_{ik} + \Phi_{ijkl} A_{jl} = 0,   \label{B8asym} 
\eea
respectively. 
We assume that our spacetime is of algebraic type II and not more special. This means that at least one of the quantities $\Phi_{ijkl}$, $\Phi_{ij}$, $\Phi$ must be non-vanishing.

In a 4d spacetime, (\ref{B4}) is trivial and (\ref{B8}) reduces to the condition that $\rho_{ij}$ is shear-free. Some consequences of these algebraic equations for higher-dimensional spacetimes were already studied in \cite{PraPraOrt07,Durkee09}. Our main objective in this section is to derive further inequivalent algebraic constraints by differentiating these equations. 

First we recall the ``Sachs equation'' \cite{OrtPraPra07,Durkeeetal10} governing the evolution of $\rho_{ij}$ along $\lb$, which for a geodesic multiple WAND reads
\begin{eqnarray}
  \tho \rho_{ij} &=& - \rho_{ik} \rho_{kj} \label{Sachs},
\end{eqnarray}
We also recall the following components of the Bianchi identity ((A10), (A11) from Ref. \cite{Durkeeetal10}):
 \begin{eqnarray}
  \tho \Phi_{ij}   &=& -(\Phi_{ik} + 2\Phia_{ik} + \Phi \del_{ik}) \rho_{kj}, \label{A2}\label{Bi2}\\[3mm]
  -\tho \Phi_{ijkl}   &=& 4\Phia_{ij} \rho_{[kl]} - 2\Phi_{[k|i} \rho_{j|l]} + 2\Phi_{[k|j} \rho_{i|l]} 
                       + 2\Phi_{ij[k|m} \rho_{m|l]} . \label{A4}\label{Bi3}
\end{eqnarray}
The idea now is to apply the derivative operator $\tho$ to \eqref{B8sym} and use these equations to eliminate derivatives from the resulting equation, to obtain a new algebraic equation\footnote{A similar procedure applied to eq. \eqref{B4} or \eqref{B8asym} does not yield a new constraint.}. After using \eqref{B4}, \eqref{B8} and $\Phi_{i[jkl]}=0$  
to simplify the result, we obtain the traceless symmetric equation
\be
\left( 2 \Phi_{kj} - \Phi_{jk} \right) \rho_{il} \rho_{jl} + \left( 2 \Phi_{ij} - \Phi_{ji} \right) \rho_{jl} \rho_{kl}  - \Phis_{ik} \rho_{jl} \rho_{jl} + \Phi \rho_{il} \rho_{kl} + \Phi_{ijkl} \rho_{js} \rho_{ls} =0. \label{thornB8}
\ee

{To summarize, in addition to  the purely algebraic equations \eqref{B4}, \eqref{B8} (this being equivalent to \eqref{B8sym} and \eqref{B8asym}) we obtained a new, purely algebraic equation \eqref{thornB8}.} 

This procedure can be repeated by acting with $\tho$ again. In this way, one obtains an infinite set of algebraic equations which are of $n$th order in $\rho_{ij}$ and linear in curvature.

\subsection{The optical constraint}
\label{Sec_OC}

Note that equation  \eqref{thornB8} is the same as equation \eqref{B8sym} with $S_{ij}$ replaced by $\rho_{ik} \rho_{jk}=(\rhob \rhob^T)_{ij}$ (where $\rhob^T$ is $\rhob$ transposed). We can write these equations as
\be
\label{Ldef}
 {\cal L}_{ijkl} S_{jl} = 0, \qquad {\cal L}_{ijkl} (\rhob \rhob^T)_{jl} = 0
\ee
where the linear map ${\cal L}$ acts on symmetric matrices and depends only on the Weyl components. The matrices $\Sb$ and $\rhob \rhob^T$ both belong to the kernel of ${\cal L}$ so this kernel must have dimension at least 1 (ignoring the trivial case $\rho_{ij}=0$). A kernel of dimension greater than 1 suggests a more constrained Weyl tensor. So a dimension 1 kernel appears to be generic. In this case, $\Sb$ and $\rhob \rhob^T$ must be proportional, which gives the {\it optical constraint} \cite{OrtPraPra09}
\be
 \rho_{ik} \rho_{jk} \propto \rho_{(ij)} . \label{eqOC}
\ee

It has been proven that the optical constraint holds for all Ricci-flat \cite{OrtPraPra09} and Einstein \cite{MalPra11GKS} Kerr-Schild spacetimes (explicit solutions within this class, such as the Myers-Perry black hole \cite{MyePer86}, are known), and for all asymptotically flat type II spacetimes admitting a non-degenerate ($\det \rhob \ne 0$) geodesic multiple WAND   \cite{OrtPraPra09b} (see eq.~(14) therein). Although the above analysis applies only to type II spacetimes, it is worth mentioning that the optical constraint also holds for all Einstein spacetimes of type N and  with additional assumptions also for  Einstein spacetimes of type III \cite{Pravdaetal04}.

In matrix notation, the optical constraint can be expressed as $\rhob \rhob^T  = \alpha \Sb $. {This implies that} $(\Id- \frac{2}{\alpha} \rhob)$ is an orthogonal matrix and therefore also  \mbox{$ \rhob^T \rhob=\alpha \Sb $}.
Since $[\rhob, \rhob^T]$ vanishes, $\rhob$ is a {\em normal} matrix and thus, using spins, can be put  to a convenient block-diagonal form which after employing the optical constraint condition reads {(see \cite{OrtPraPra09,OrtPraPra10} for  related discussions)} 
\begin{small}
\bea
\rhob =\alpha\  {\rm diag}\left(1 , \dots, 1 , 
\frac{1}{1+ \alpha^2 b_1^2}\left[\begin {array}{cc} 1 & -\alpha b_1 \\ 
   \alpha b_1 & 1  \label{canformL} \\
  \end {array}
 \right]
, \dots, 
\frac{1}{1+ \alpha^2 b_\nu^2} \left[\begin {array}{cc} 1 & -\alpha b_\nu \\ 
   \alpha b_\nu & 1 \\ \end {array}
 \right] 
 , 0, \dots ,0
\right).
\eea
\end{small}
Note that the symmetric part of each 2-block is proportional to a 2-dimensional identity matrix, i.e. it is ``shear-free''. 

In 4d, the optical constraint implies that either (i) there is a single $2 \times 2$ block, in which case $\rho_{ij}$ is shearfree; or (ii) $\rho_{ij}$ is symmetric with exactly one non-vanishing eigenvalue. However, the Goldberg-Sachs theorem shows that case (ii) cannot occur. So in 4d the optical constraint is a necessary condition for $\lb$ to be a repeated principal null direction, but it is not sharp.

In 5d, the optical constraint implies that an orthonormal basis can be found for which the optical matrix takes one of the forms \eqref{5dform1}, \eqref{5dform2} or, with $a=0$, \eqref{5dform3} of Theorem 1. The form \eqref{5dform3} with $a,b\ne 0$ violates the optical constraint. However, as we will discuss later, known solutions with an optical matrix of the latter form are of type D and  admit another geodesic multiple WAND which has an optical matrix of the form \eqref{5dform3} with $a=0$, and therefore satisfies the optical constraint. It is thus an open question whether 
genuine type II spacetimes with optical matrix of the form \eqref{5dform3} with $a\ne 0$  exist.

We emphasize that we will {\it not} be assuming the optical constraint in what follows.

\section{Type II spacetimes in five dimensions}

\label{sec_5D}

\subsection{Algebraic constraints in five dimensions}

In five dimensions the above equations can be considerably simplified since the $\Phi_{ijkl}$ components are determined by $\Phis_{ij}$ via \cite{PraPraOrt07}
 \be
 \Phi_{ijkl} = 2 \left(\delta_{il} \Phis_{jk} -  \delta_{ik} \Phis_{jl} -
\delta_{jl} \Phis_{ik} + \delta_{jk} \Phis_{il}  \right) - \Phi \left(
\delta_{il} \delta_{jk} - \delta_{ik} \delta_{jl}   \right). \label{Weyl5D} 
\ee
For the spacetime to be type II and not more special, we must have $\Phi_{ij} \ne 0$. Equations \eqref{B8sym}, \eqref{B8asym} reduce to 
\bea
\Phi_{ij} S_{jk} + \Phi_{kj} S_{ji} - \rho \Phis_{ik} = \frac{1}{3}  \delta_{ik} \left( 2 \Phi_{jl} S_{jl} -\rho \Phi   \right), \label{B8sym5D}  \\
2 \Phi_{jk} A_{ji} + 2 \Phi_{ji} A_{kj} + \Phi_{ij} S_{jk} - \Phi_{kj} S_{ji} + \rho \Phia_{ki} + 2 \Phi A_{ik} =0,  \label{B8asym5D}
\eea
respectively, while \eqref{thornB8} becomes
\be
\Phi_{ij} \rho_{jl} \rho_{kl} + \Phi_{kj} \rho_{jl} \rho_{il} - \rho_{jl} \rho_{jl} \Phis_{ik} = \frac{1}{3}  \delta_{ik} \left( 2 \Phi_{jl} \rho_{js}\rho_{ls}   -\rho_{jl} \rho_{jl}  \Phi   \right). \label{thornB85D} \\
\ee
Note that in five dimensions  \eqref{B4} is equivalent to \eqref{B8asym5D}.

Since \eqref{B8sym5D} is traceless, equations \eqref{B8sym5D} and \eqref{B8asym5D} can be viewed as eight linear equations for the nine unknown components of $\Phi_{ij}$. Equation \eqref{thornB85D} is traceless and symmetric and, in general, gives an additional five linear equations and thus the system \eqref{B8sym5D} -- \eqref{thornB85D} is, for general $\rhob$, overdetermined. In this section we  determine all possible forms of $\rhob$ for which there exist non-vanishing solutions of \eqref{B8sym5D} -- \eqref{thornB85D}.

To simplify the following calculations we choose the spatial basis vectors $\mb{i}$ so that $\Sb$ (the symmetric part of $\rhob$) is diagonal:
\be
 \Sb = {\rm diag}(s_2, s_3,s_4)
\ee
Due to the tracelessness of  \eqref{B8sym5D} we  replace the three
diagonal components of \eqref{B8sym5D} by their linear combinations
\bea
({s_3}-{s_2}-{s_4})  \Phis_{33} -({s_4}-{s_2}-{s_3})  \Phis_{44} =0,  \label{hkomb23} \\
({s_2}-{s_4}-{s_3})  \Phis_{22} -({s_4}-{s_2}-{s_3})  \Phis_{44} =0,  \label{hkomb34}
\eea
It is convenient to divide then the analysis into two cases. The first case corresponds to a non-twisting multiple WAND, i.e. $A_{ij}=0$. The second case is $A_{ij} \ne 0$.

\subsection{Non-twisting case ($A_{ij}=0$)}
\label{sec_non_twist}

\subsubsection{Non-twisting case with $\Phia_{ij} \not= 0$.}

Let us first study the non-twisting case with $\Phia_{ij} \not= 0$. Without loss of generality we can assume $\Phia_{23} \not= 0$. The $i=2$, $k=3$ components of 
\eqref{B8sym5D}, \eqref{B8asym5D} and  \eqref{thornB85D} read
\BEA
                    \left( s_{3}  - s_{2} \right) \Phia_{23} - s_{4} \Phis_{23}  = 0, \\
                    \left( s_{3}  - s_{2} \right) \Phis_{23} - s_{4} \Phia_{23}  = 0, \\
                   \left( {s_{3}}^2 -{s_{2}}^2 \right) \Phia_{23}- {s_{4}}^2 \Phis_{23}  = 0,
\EEA
respectively. 
This is a system of three linear equations for two unknowns $\Phis_{23}$ and $\Phia_{23}$. Thus in order to have non-vanishing $\Phia_{23}$, all determinants corresponding to pairs of these equations have to vanish.  Two such determinants give
\BEA
{s_4}^2-(s_{3}-s_{2})^2 = 0,\\
{s_4} \left(s_{2}-s_{3}\right)\left(s_{2}+s_{3}-s_4\right)=0.
\EEA
It follows that {\em for the non-twisting case with $\Phia_{ij} \not= 0$,  $\Sb$  possesses a pair of equal eigenvalues with the remaining eigenvalue being zero} 
(cf. also Proposition~10 of \cite{PraPraOrt07} and Proposition~7 of \cite{Durkee09}). 
In Appendix~\ref{app}, we prove that this result generalizes to any number of dimensions.

\subsubsection{Non-twisting case with  $\Phia_{ij} = 0$.}
\label{Sec_nontwistPHIA0}

In the non-twisting case with $\Phia_{ij} = 0$,  eq. \eqref{B8asym5D} implies that $\Phi_{ij}$ and $S_{ij}$ commute:
\be
[\Sb,\Phib]=0.
\ee
Hence we can choose our spatial basis vectors so that both $\Phi_{ij}$ and $S_{ij}$ are diagonal ($\Phi^A_{ij}=0$ implies that $\Phi_{ij}$ is symmetric).
Equation \eqref{thornB85D} reduces to 
\bea
({s_3}^2-{s_2}^2-{s_4}^2)  \Phis_{33} -({s_4}^2-{s_2}^2-{s_3}^2)  \Phis_{44} =0,  \label{dhkomb23} \\
({s_2}^2-{s_4}^2-{s_3}^2)  \Phis_{22} -({s_4}^2-{s_2}^2-{s_3}^2)  \Phis_{44} =0.  \label{dhkomb34}
\eea
Equations \eqref{hkomb23}, \eqref{hkomb34}, \eqref{dhkomb23} and \eqref{dhkomb34} form a set of four linear equations for three unknowns
$\Phis_{22}$, $\Phis_{33}$, $\Phis_{44}$. Thus in order to allow for non-vanishing $\Phi_{ij}$, the determinant of each triplet of these equations has to vanish. Two such determinants give
\BEA
{s_2}  \left({s_3}-{s_4}\right) \left(-{s_3}-{s_4}+{s_2} \right)^2 =0,   \\
{s_3} \left({s_2}-{s_4}\right) \left(-{s_3}+{s_2}+{s_4}\right)^2 =0,
\EEA
which implies that all non-vanishing eigenvalues of  $\rho_{ij}$ coincide. 

\subsubsection{Summary of non-twisting case}

\label{subsucsec_nontwist_summary}

Together with the results for types N and III obtained\footnote{To be precise, Ref.~\cite{Pravdaetal04} considers only Ricci flat spacetimes, however { such results extend immediately to Einstein spacetimes since the Ricci tensor does not appear in the Bianchi identity in that case.}} in \cite{Pravdaetal04} our results can be summarized as
\begin{prop}
In a five-dimensional Einstein spacetime admitting a non-twisting (and thus geodesic) multiple WAND, all non-vanishing eigenvalues of $\rho_{ij}$ coincide. 
\end{prop}
This leads to the following cases:

\vspace{2mm}\noindent
{\it Rank 3 optical matrix:} This is case i) of Theorem~\ref{th_GS} with $a=0$. These solutions constitute the Robinson-Trautman class, their metrics are explicitly known \cite{PodOrt06} and in five dimensions they reduce to the Schwarzschild-Tangherlini metric with a possible cosmological constant. 
{We observe that, apart from Kundt solutions, these are the only 5d Einstein spacetimes admitting a geodesic {\em shearfree} multiple WAND (see Section~\ref{subsubsec_shearfree} below).}
 From \eqref{hkomb23}, \eqref{hkomb34}, $\Phi_{ij}=\frac{\Phi}{3}\delta_{ij}$, which fully determines the (type D \cite{PodOrt06}) Weyl tensor. This corresponds to the spin type $\{(000)\}_0[\Phi\neq0]$ in the refined five-dimensional classification recently proposed in \cite{Coleyetal12}.\footnote{Note that our $\Phi$ is essentially the quantity $\bar R$ of \cite{Coleyetal12} (see Table~1 therein).}

\vspace{2mm} \noindent
{\it Rank  2 optical matrix:} This is case ii) of Theorem~\ref{th_GS} with $a=0$. In principle, this case can have $\Phia_{ij} \not=0$, however, we are not aware of explicit examples in this class. As in the twisting case, $\Phi_{ij}$ is constrained by eq.~(\ref{Phi_ii}) below\footnote{We have checked that this equation holds in the non-twisting case as well.}
 (i.e., $\Phi_{i4}=0$) and the spin type is, in general, $\{111\}_{\mbox g}[\Phi\neq0]$.
A direct or warped product of any 4d Einstein type II  Robinson-Trautman metric leads to a five-dimensional Einstein type II metric with optical matrix of this type \cite{OrtPraPra11} (and the spin type specializes to $\{(11)1\}_0[\Phi\neq0]$ in this case). This includes the Schwarzschild black string solution.

\vspace{2mm} \noindent
{\it Rank 1 optical matrix:} This is case iii) of Theorem~\ref{th_GS} with $a=0$. (It can also be obtained as a limit of case i): $a\rightarrow \infty$, $b\rightarrow 0$, $a^2 b$ fixed.) From \eqref{hkomb23}, \eqref{hkomb34}, $\Phi_{ij} = {\rm diag}(-\Phi,\Phi,\Phi)$, and the spin type is  $\{(11)1\}_0[\Phi\neq0]$. 
An example {(of type D)} belonging to this class is a {non-twisting, expanding and shearing} geodesic multiple WAND in $dS_3 \times S^2$ {(see Section~\ref{subsubsec_ex_iii}).} 
Another example  of type D is the 
Kaluza-Klein bubble solution \cite{Witten82} (analytically continued 5d Schwarzschild) discussed in Ref. \cite{DurRea09}.

\vspace{2mm} \noindent
{\it Vanishing optical matrix:} This is any of the cases of Theorem~\ref{th_GS} with $b=0$. This is the Kundt class.  

\subsection{Twisting case ($A_{ij} \ne 0$)}

\subsubsection{Determinants}

Now, let us  study the twisting case $A_{ij} \ne 0$. Let $D_{ij}$ be the determinant of the system of linear equations consisting of  \eqref{hkomb23}, \eqref{hkomb34}, {the} three off-diagonal components of \eqref{B8sym5D}, {the} three components of {\eqref{B8asym5D} } and {the} $i,j$ component of \eqref{thornB85D}. If the system of all algebraic constraints obtained above is to admit a non-vanishing solution $\Phi_{ij}$ then all  $D_{ij}$ must vanish. This    considerably restricts the possible forms of $\rhob$.

{For $i\neq j$, one obtains}, up to {overall} numerical factors, 
\bea
D_{23} &=& \left[ 2 {s_4} ({s_3} - {s_2}) ({s_4}-{s_2}-{s_3}) A_{23} - A_{34}A_{24}({s_2} - {s_3} - {s_4}) ({s_3} - {s_2} - {s_4})  \right] F,   \label{eqD23} \\
D_{24} &=& \left[ 2   {s_3} ({s_4}-{s_2}) ({s_3}-{s_2}-{s_4})A_{24}  + A_{34}A_{23}  ({s_2}-{s_3}-{s_4}) ({s_4}-{s_2}-{s_3})   \right] F, \label{eqD24} \\
D_{34} &=& \left[ 2 {s_2} ({s_4}-{s_3}) ({s_2}-{s_3}-{s_4})  A_{34} - A_{23}A_{24}  ({s_4}-{s_2}-{s_3}) ({s_3}-{s_2}-{s_4}) \right] F,\label{eqD34}
\eea
where 
\bea
F &=& ({s_4}-{s_2}-{s_3})^2({s_3}-{s_2}-{s_4})^2 ({s_2}-{s_3}-{s_4})^2+4[  A_{34} ^2 ({s_4}-{s_2}-{s_3})^2({s_3}-{s_2}-{s_4})^2
\nonumber 
\\ & &+  A_{24} ^2 ({s_4}-{s_2}-{s_3})^2({s_2}-{s_3}-{s_4})^2 
+  A_{23} ^2 ({s_2}-{s_3}-{s_4})^2 ({s_3}-{s_2}-{s_4})^2 ]. \label{eqF} 
\eea
The structure of the above determinants suggests  that cases  with $s_i+s_j=s_k$ or $s_i=s_k$ or $s_i=0$ for some distinct values of $i,j,k$ should be studied separately. Therefore  first we study the ``generic'' case with $s_i+s_j\not=s_k$ and $s_i\not=s_k$, $s_i\not=0$ for all distinct values of $i$, $j$, $k$.  

\subsubsection{The case with $s_i+s_j\not=s_k$ and $s_i\not=s_k$, $s_i\not=0$ for  distinct values of $i$, $j$, $k$.}
\label{Sec_gen}
With this assumption,
$F$ is non-vanishing. Now it follows from eqs. \eqref{eqD23} - \eqref{eqD34} that vanishing of one component of $A_{ij}$ implies vanishing of all $A_{ij}$. Since here we consider non-zero twist, all components of $A_{ij}$ are {necessarily} non-zero. 

Now expressing $A_{23}$ from (\ref{eqD23}),
Eqs. (\ref{eqD24}), (\ref{eqD34})  reduce to
\bea
   A_{34} ^2 ({s_2}-{s_3}-{s_4})^2+4 {s_3} ({s_4}-{s_2}) {s_4} ({s_3}-{s_2}) = 0, \label{eqA34}\\
   A_{24} ^2 ({s_3}-{s_2}-{s_4})^2  - 4 {s_2} ({s_4}-{s_3}) {s_4} ({s_3}-{s_2}) = 0 \label{eqA24},
 \eea
 respectively.

Since $D_{22}$, $D_{33}$ and $D_{44}$ contain only squares of components of $A_{ij}$, using (\ref{eqD23}), (\ref{eqA34}) and (\ref{eqA24}),  we can express them in terms of $S_{ij}$ only. We find that
\be
\frac{D_{22}-D_{33}}{{s_4} (s_2-{s_3})} - \frac{D_{22}-D_{44}}{{s_3} (s_2-{s_4})} = k ({s_3}-{s_4}) s_2,
\ee
where $k$ is a non-zero numerical constant. By assumption, the RHS is non-zero hence not all of the determinants $D_{ij}$ can vanish. Therefore {\em type II Einstein spacetimes with $\rhob$ obeying $s_i+s_j\not=s_k$ and $s_i\not=s_k$, $s_i\not=0$ (for  distinct values of $i$, $j$, $k$) do not exist}. 

\subsubsection{Case with all  $s_i$ non-vanishing and distinct, with $s_4 = s_3 + s_2$}
\label{sec_case2}

Let us now study the branch with $s_i = s_j + s_k$ for some values of $i,j,k$. Without loss of generality
we can then assume $s_4 = s_3 + s_2$. With this assumption, the vanishing of $D_{24}$ and $D_{34}$ implies
\bea
D_{24} &=& 0 \ \Rightarrow  \   {A_{23}} A_{24} {s_{2}} {s_{3}} = 0, \\ 
D_{34} &=& 0 \ \Rightarrow \   {A_{23}} {A_{34}} {s_2} {s_3} = 0. 
\eea
Thus either $A_{23}=0$ or $A_{24}=A_{34}=0$.\\

\noindent {\em Subcase $A_{23}=0$: }
In this subcase $F=0$  and consequently all determinants $D_{ij}$ vanish. It is thus necessary to study various sets of equations containing more than one equation
from (\ref{thornB85D}). In particular,  vanishing of determinant of the system [(\ref{hkomb23}), off-diagonal components of \eqref{B8sym5D}, [2,4], [3,4] components of \eqref{B8asym5D}, [2,2], [2,4], [3,3] components
of (\ref{thornB85D})] implies
\be
{s_3}^3 {s_2}^4 {A_{34}}^3 ({A_{24}^2}+{s_2}^2) = 0  \ \ \Rightarrow  \ \ A_{34}=0, \label{syst1}
\ee
while vanishing of determinant of the system [(\ref{hkomb34}), off-diagonal components of \eqref{B8sym5D}, [2,4], [3,4] components of \eqref{B8asym5D}, [2,2], [2,3], [4,4] components
of (\ref{thornB85D})] implies
\be
{s_3}^4 {s_2}^2 {A_{24}}^2 ({A_{34}}^2 {s_2}^2 + {s_3}^2 {A_{24}}^2) = 0 \Rightarrow  \ \ A_{24}=0 , \label{syst2}
\ee 
{and therefore $A_{ij}=0$, which is a contradiction.}

\noindent {\em{Subcase $A_{24}=A_{34}=0$:}}
In this subcase vanishing of $D_{33}$ and $D_{44}$ imply
\bea
(2 {s_2}^2-{s_3}^2) {A_{23}}^2-{s_2}^2 {s_3}^2 = 0,\\
({s_2}^2+{s_3}^2) {A_{23}}^2-2 {s_2}^2 {s_3}^2 = 0,
\eea
respectively. However, this  also implies ${s_2}^2 = {s_3}^2$, which is not compatible with the assumptions of this section (all  $s_i$ non-vanishing and distinct with $s_4 = s_3 + s_2$).

Putting together these subcases we conclude that {\em there are no twisting Einstein type II solutions obeying assumptions of this section}. Together with the results of Sections \ref{sec_non_twist} and \ref{Sec_gen} this implies
\begin{prop}
For a geodesic multiple WAND in a  five-dimensional type II Einstein spacetime,
  at least two eigenvalues  of $\rho_{(ij)}$ coincide or at least one eigenvalue vanishes.
\label{prop_5Dtwist}
\end{prop}
The rest of the analysis is based on studying all cases compatible with this proposition. 

\subsubsection{Case with all eigenvalues $s_i$ non-vanishing with $s_2=s_3 \not= s_4$ }

\label{subsubsec_5D_3}

Now $D_{23}=0$ implies
\be
 A_{24} A_{34}  {s_4}    \left[({s_4}-2 {s_2})^2(4A_{34}^2+4A_{24} ^2+{s_4}^2)+4  A_{23} ^2\right] = 0.
\ee
Therefore we have two cases to consider:\\
a) $A_{23}=0,\ s_{4}=2 s_{2}$; \\ 
b) $A_{24} A_{34}=0$. Without loss of generality we can set $A_{24}=0$.\\

\noindent
{\it Case a)} Vanishing of the determinant of the systems used in \eqref{syst1} and \eqref{syst2} implies
\bea
A_{34} ({s_{2}}^2+{A_{24}}^2) = 0,\\
A_{24} ({A_{34}}^2 {s_{2}}^2+{s_{3}}^2 {A_{24}}^2) = 0,
\eea
respectively, which yields $A_{24}=A_{34}=0$. Hence $A_{ij}=0$, which is a contradiction.

\noindent
{\it Case b)}  Now $D_{22}-D_{33}=0$  implies $A_{34}=0$ and then $D_{33}=0$ leads to
\be
s_4 = \frac{s_2^2+A_{23}^2}{s_2} \label{nondeg_MP}.
\ee

We thus arrived to $\rhob$ of the form 
\be
S_{ij}= \left( \begin {array}{ccc} s_2&0&0
\\\noalign{\medskip}0& s_2&0
\\\noalign{\medskip}0&0&s_4\end {array}
 \right),
\ \ \ \
A_{ij}=    \left( \begin {array}{ccc}
0&A_{23}&0\\
-A_{23}&0&0\\
0&0&0\end {array}
 \right)  ,
\ee
 where $s_4$ is given by (\ref{nondeg_MP}).   This is case i) of Theorem~\ref{th_GS}.

After some algebra, the general solution of \eqref{B8sym5D} -- \eqref{thornB85D} is
\be
 \Phis_{ij} =  \left( \begin {array}{ccc} \Phis_{22}&0&0
\\\noalign{\medskip}0& \Phis_{22} &0
\\\noalign{\medskip}0&0&\Phis_{44}\end {array}
 \right),
\ \ \ \
\Phia_{ij} =   \left( \begin {array}{ccc}
0&\Phia_{23}&0 \\
-\Phia_{23}&0&0\\
0&0&0\end {array}
 \right), 
 \ee
where the Weyl components $\Phis_{22}$ and $\Phia_{23}$ are given by 
\be
  \Phis_{22}  = \frac{{s_2}^2- A_{23} ^2 }{  {s_2}^2+ A_{23} ^2} \Phis_{44},  \qquad \Phia_{23}  = \frac{2 A_{23}  {s_2}}{{s_2}^2+ A_{23} ^2}  \Phis_{44}.  \label{WeylMP}
\ee

Note that in this case $[\rhob,\Phib]=0$, $\rhob$ obeys the optical constraint and $\Phis_{44}$ has to be non-vanishing (and thus also $\Phia_{23}\neq 0$, as long as there is twist). The spin type is $\{(11)1\}_\parallel[\Phi\neq0]$. This class of solutions contains, e.g., the Myers-Perry black holes (cf.~\cite{PraPraOrt07}), and KK bubbles discussed in Section \ref{sec_form_i}.  Note that the non-twisting limit can lead either to a rank 3 optical matrix (this holds e.g. for Myers-Perry black hole) or rank 1 optical matrix (when $a\rightarrow \infty$, $b\rightarrow 0$, with $a^2 b$ fixed and non-vanishing - this holds for KK bubbles).

\subsubsection{Shearfree case $s_2=s_3=s_4$}

\label{subsubsec_shearfree}

Since we are in odd dimensions, by Proposition~3 from Ref. \cite{OrtPraPra07} the shearfree case is non-twisting, which contradicts the assumption of this section.

\subsubsection{Case $\mbox{rank}(\Sb)=2$, $s_2 \not=s_3$, $s_4=0$ }
\label{Sec_nonOC}

First let us discuss the case with vanishing $F$: this holds iff $s_2+s_3=0=A_{23}$ (see eq. \eqref{eqF}).
Then the systems used in \eqref{syst1} and
\eqref{syst2} imply
\bea
{s_2} A_{34}  ({s_2}^2+{A_{24}}^2) =0, \\
{s_2} {A_{24}} ({A_{34}}^2+{A_{24}}^2) = 0,
\eea 
respectively and therefore $A_{24}=A_{34}=0$, {i.e., $A_{ij}=0$}, which is a contradiction. In the rest of this section we can thus assume that $F \ne 0$. Now $D_{23}=0$ and $D_{22} - D_{33} =0$ imply 
\bea
A_{24} A_{34} ({s_2}-{s_3})^2 = 0, \\
 ({A_{34}}^2-{A_{24}}^2) ({s_2}-{s_3})^2 = 0,
\eea
and therefore  $A_{24}=A_{34}=0$. $D_{22}=0$ then implies 
\be
 ({s_2}-{s_3})  ({s_2} {s_3}+ {A_{23}}^2) = 0
\ee
and therefore
\be
{s_3} = -\frac{A_{23}^2}{{s_2}} \label{cond_s3deg}.
\ee
We have shown that $\rhob$ has the form
\be
S_{ij}= \left( \begin {array}{ccc} s_2&0&0
\\\noalign{\medskip}0& s_3&0
\\\noalign{\medskip}0&0&0\end {array}
 \right),
\ \ \ \
A_{ij}=    \left( \begin {array}{ccc}
0&A_{23}&0\\
-A_{23}&0&0\\
0&0&0\end {array}
 \right)  ,
\ee
where $s_3$ is given by \eqref{cond_s3deg}. Note that $\rhob$ has rank 1. This is case iii) of Theorem~\ref{th_GS} with $a \ne 0$, for which the optical constraint is violated ($\rhob$ is not even a normal matrix). For explicit examples see Section \ref{subsubsec_ex_iii}. 

In this case, the solution of the system of equations \eqref{B8sym5D} -- \eqref{thornB85D} is 
\be
\label{PhinonOC}
 \Phi_{ij} =  \left( \begin {array}{ccc} \Phis_{22}&\Phis_{23}&0
\\\noalign{\medskip}\Phis_{23}& -\Phis_{22} &0
\\\noalign{\medskip}0&0&\Phi\end {array}
 \right),
\ee
where
\be
 \Phis_{22}  = \frac{ A_{23}^2-{s_2}^2 }{{s_2}^2+A_{23}^2} \Phi,   \qquad   \Phis_{23} = \frac{2 {s_2} A_{23}}{{s_2}^2+A_{23}^2} \Phi.   \label{eqWeylrank2}
\ee
Note that  in this case $\Phia_{ij}=0$. Substituting these results into \eqref{A2} gives the simple result
\be
 \tho \Phi_{ij} = 0.
\ee
If one uses a basis parallelly transported along a geodesic tangent to $\lb$ then this equation implies that the components of $\Phi_{ij}$ are constant along the geodesic.

If one performs a change of basis to diagonalize $\Phi_{ij}$ then the result is $\Phi_{ij} = {\rm diag}(-\Phi,\Phi,\Phi)$ (which also shows that the spin type is  $\{(11)1\}_0[\Phi\neq0]$). {This observation will be useful in the following.}

\subsubsection{Case $\mbox{rank}(\Sb)=2$, $s_2=s_3\neq0$, $s_4=0$ }

\label{subsubsec_stringtype}

For this class, vanishing of determinant of a system consisting of
[\eqref{hkomb23}, \eqref{hkomb34}, [2,4] and [3,4] components of  \eqref{B8sym5D}, \eqref{B8asym5D}, [2,2], [2,4] and [3,4] components of  \eqref{thornB85D}]
leads to
\be
 {s_2} \left({A_{24}}^2+{A_{34}}^2\right)=0,
\ee
and therefore ${A_{24}}={A_{34}}=0$. Hence $\rhob$ has the form 
\be
S_{ij}= \left( \begin {array}{ccc} s_2&0&0
\\\noalign{\medskip}0& s_2&0
\\\noalign{\medskip}0&0&0\end {array}
 \right),
\ \ \ \
A_{ij}=    \left( \begin {array}{ccc}
0&A_{23}&0\\
-A_{23}&0&0\\
0&0&0\end {array}
 \right)  .
\ee
This class represents case ii) of Theorem~\ref{th_GS} with $a \ne 0$.  Note that $\rhob$ satisfies the optical constraint. An example of a solution belonging to this class is the Kerr black string. {See Section~\ref{subsubsec_ex_ii} for more examples.}

For this case, the system of equations \eqref{B8sym5D} -- \eqref{thornB85D} reduces to
\be
\Phia_{24}=-\Phis_{24}, 
\quad \Phia_{34}= - \Phis_{34}, \quad \Phis_{44}=0.
 \label{Phi_ii}
\ee
This means that $\Phi_{ij}$ has rank 2. The spin type is in general $\{111\}_{\mbox g}[\Phi\neq0]$ (but it simplifies to $\{(11)1\}_\parallel[\Phi\neq0]$ for, e.g., the Kerr black string).

\subsubsection{Case $\mbox{rank}(\Sb)=1$, $s_2\neq0$, $s_3=s_4=0$}

In this case $D_{23}=0$, $D_{24}=0$, $D_{34}=0$ imply
\be
A_{34}A_{24} = 0, \ \
A_{34}A_{23} = 0, \ \
A_{23}A_{24} = 0,
\ee
respectively, and $D_{22}=0$, $D_{33}=0$, $D_{44}=0$ give
\bea
-2{A_{34}}^2+{A_{23}}^2+{A_{24}^2} = 0,\\
{A_{34}}^2+{A_{23}}^2-2{A_{24}^2} = 0,\\
-{A_{34}}^2+2{A_{23}}^2-{A_{24}^2} = 0.
\eea
Therefore there are no twisting solutions belonging to this class.

\subsubsection{Case with vanishing $S_{ij}$}

By Proposition 1 of \cite{OrtPraPra07} $A_{ij}=0$, which is a contradiction.

\subsection{Summary}

We have considered all possible cases and demonstrated that they obey Theorem~\ref{th_GS}, so we have proved the theorem.

\subsection{Type D spacetimes}

In the case of type D spacetimes we can extract some more information from the above general results.

\subsubsection{Type D Einstein spacetimes violating the optical constraint}

\label{sec_D_iii}

Consider a type D spacetime with multiple WANDs $\lb$, $\nb$ with $\lb$ geodesic. Assume that $\lb$ violates the optical constraint, i.e., it corresponds to case (iii) of Theorem~\ref{th_GS} with $a,b \ne 0$. In deriving the above results, the only basis transformations used were rotations of the spatial basis vectors $\mb{i}$ and so $\nb$ remains a multiple WAND. In Section~\ref{Sec_nonOC}, we mentioned that there is a basis in which $\Phi_{ij} = {\rm diag}(-\Phi,\Phi,\Phi)$. For a type D solution, one can use this to repeat the argument in section 4 of Ref. \cite{DurRea09} to show that at any point $\lb$ belongs to a $1$-parameter family of multiple WANDs. This parameter becomes a free function in spacetime. In general, this function will not satisfy the PDE required for the corresponding multiple WAND to be geodesic. Hence we have shown 
that
all five-dimensional type D Einstein spacetimes violating the optical constraint admit a non-geodesic multiple WAND.

All such solutions were classified in Ref. \cite{DurRea09}, and they are the solutions listed {in Section~\ref{subsubsec_ex_iii} below}. In fact type D Einstein spacetimes admitting a non-geodesic multiple WAND admit a canonical form
$\Phi_{ij} = {\rm diag}(-\Phi,\Phi,\Phi)${\cite{PraPraOrt07}} which is not compatible with cases (i) and (ii) of  Theorem~\ref{th_GS}.
Thus in the type D case all solutions admitting a geodesic multiple WAND which violates the optical constraint are  explicitly known, and {\em coincide}\footnote{For all type D metrics admitting non-geodesic multiple WANDs we present examples of geodetic WANDs not obeying the optical constraint in Section~\ref{subsubsec_ex_iii}.} 
with the family of Einstein spacetimes admitting a non-geodesic multiple WAND \cite{DurRea09}:

\begin{prop}
A five-dimensional type D Einstein spacetime admits a geodesic multiple WAND violating the optical constraint if and only if it admits a non-geodesic multiple WAND.
\label{prop_nongeod}
\end{prop}

Note that all such spacetimes also admit non-twisting geodesic multiple WANDs which respect the optical constraint, with optical matrix of rank 1 {(namely, when $\alpha$ is a constant in Section~\ref{subsubsec_ex_iii})}.
In other words, in any 5d type D Einstein spacetime, one can choose two distinct multiple WANDs which respect the optical constraint. It is conceivable that  
genuine type II spacetimes with an optical matrix that does not satisfy the optical constraint do not exist.
If so, we would have the result that a 5d Einstein spacetime admits a geodesic multiple WAND if and only if it admits a geodesic multiple WAND that satisfies the optical constraint. 

\subsubsection{Possible optical properties of the multiple WANDs}

Type D spacetimes admit (at least) two multiple WANDs. Without loss of generality, we can always consider both of them  to be 
geodesic\footnote{If one of the two multiple WANDs is non-geodesic then the spacetime belongs to the case (iii) of Theorem~\ref{th_GS} and thus also contains more than one multiple geodesic multiple WANDs, e.g. for different choices of the parameter $\alpha$ in \eqref{l_S2_AdS3} or \eqref{l_Schw}.} \cite{DurRea09} and identify those with the frame vectors $\lb$ and $\nb$. The results of Theorem~\ref{th_GS} thus apply to the optical matrices $\rhob$ and  $\rhob'$ of both multiple WANDs (although, in general, the canonical frames will be different for each of them). We want to discuss here what are the permitted combinations of those types for $\rhob$ and  $\rhob'$, respectively, in the case when $\rhob\neq 0\neq\rhob'$. This can be done by looking at invariant properties of the matrix $\Phi_{ij}$ associated with the various cases (since interchanging $\lb$ and $\nb$ does not affect $\Phi_{ij}$, except for changing the sign of $\Phi^A_{ij}$), as discussed in Section~\ref{subsucsec_nontwist_summary} for the non-twisting case, and in Sections~\ref{subsubsec_5D_3}, \ref{Sec_nonOC} and \ref{subsubsec_stringtype} for the twisting case.

First, note that when $\rhob$ is of type (ii) then $\Phi_{ij}$ has rank 2 while $\Phi_{ij}$ has always rank 3 in the remaining cases. Therefore if $\rhob$ is of type (ii) then also $\rhob'$ must necessarily be such. Next, in case~(i) either $\Phi^A_{ij}\neq 0$ or  $\Phi_{ij}\sim\delta_{ij}$ and neither of these is compatible with the form of $\Phi_{ij}$ permitted in case (iii). Therefore it is not possible that $\rhob$ is of type (i) while $\rhob'$ being of type (iii). Hence we have shown that in type D Einstein spacetimes if two multiple WANDs have both non-zero optical matrices then these must fall into the same case (i), (ii) or (iii) of Theorem~\ref{th_GS}. Note, in addition, that if $\rhob$ and $\rhob'$ are both of type (i) then they must be either both twisting (if $\Phi^A_{ij}\neq 0$) or both non-twisting (if $\Phi^A_{ij}=0$, which is the Robinson-Trautman case).\footnote{In four dimensions, a corresponding result says that diverging PNDs of Einstein spacetimes of type D are twist-free if and only if the Weyl tensor is purely electric \cite{McIntoshetal94,WyllBeke10}.} On the other hand, the examples of Section~\ref{subsubsec_ex_iii} show that a non-twisting $\rhob$ of type (iii) can pair both with a non-twisting and with a twisting $\rhob'$ of type (iii). 
In fact, let us note that the (fully known) case (iii) defines the class of five-dimensional Einstein spacetimes that admit more than two multiple WANDs,\footnote{The set of multiple WANDs is actually infinite and homeomorphic to a 1-sphere \cite{ParWyl11,Wylleman12} (some of these results were given already in \cite{DurRea09}).} whereas in cases (i) and (ii) there exist precisely two multiple WANDs.

\section{Existence of an optical structure}

\label{sec_integrab}

\subsection{Definition of ``optical structure'' \cite{Taghavi-Chabert11}}

In four dimensions, the existence of a null congruence $\lb$ that is geodesic and shearfree (g.s.) can be rephrased in terms of properties of different geometric objects, leading to various formulations of the g.s. condition (see, e.g., \cite{penrosebook2,RobTra83,NurTra02,Trautman02a,Trautman02b}). Accordingly, the Goldberg-Sachs theorem can be equivalently rephrased in terms of any of such geometric properties. The condition of $\lb$ being geodetic and shearfree can be replaced, in particular, by a statement about the existence of an {\em integrable} complex two-dimensional totally null distribution (defined by the span of $\lb$ and $\mb{2}+i\mb{3}$). It turns out that in higher dimensions, those various formulations of the g.s. property are not equivalent and one could thus expect, in principle, that the Goldberg-Sachs theorem might admit several inequivalent extensions in more than four dimensions. A possible five-dimensional version of the Goldberg-Sachs theorem has been proposed in \cite{Taghavi-Chabert11}, where the notion of a g.s. null congruence has been replaced by that of an ``optical structure''. Here we discuss consequence of our results (in particular, Theorem~\ref{th_GS}) in terms of existence of such an optical structure, and compare our conclusions with those of \cite{Taghavi-Chabert11}.

Let us define the totally null distribution
\be
 {\cal D}=\mbox{Span}\{\mb{2}+i\mb{3},\bl\} ,
 \label{D}
\ee
and its orthogonal complement 
\be
 {\cal D}^\bot=\mbox{Span}\{\mb{2}+i\mb{3},\mb{4},\bl\} .
 \label{D_orth}
\ee

By definition \cite{Taghavi-Chabert11}, there is an {\em optical (or Robinson) structure} on the five-dimensional spacetime if 
both ${\cal D}$ and ${\cal D}^\bot$ are integrable, which is equivalent to
\beqn
 & & \kappa_{i}=0, \qquad \rho_{33}=\rho_{22}, \qquad \rho_{32}=-\rho_{23},\qquad \rho_{24}=0=\rho_{34}, \qquad \rho_{42}=0=\rho_{43},  			  \label{Dorth_integr_k_rho} 
 \\  & & \M{2}{4}{0}=0=\M{3}{4}{0} , \qquad \M{2}{4}{2}=\M{3}{4}{3}, \qquad \M{2}{4}{3}=-\M{3}{4}{2} .
 \label{Dorth_integr_M} 
\eeqn
See \cite{Pravdaetal04,OrtPraPra07,Durkeeetal10} for the definition of $\M{a}{b}{c}$.

Obviously if an optical structure is integrable its complex conjugate is integrable too, and this will be understood in the following.

\subsection{Sufficient conditions for the existence of an optical structure}

The following proposition provides sufficient conditions for the existence of an optical structure in algebraically special Einstein spacetimes in five dimensions. Note that we need the assumption that the optical matrix of a multiple WAND is non-zero, but on the other hand the conditions on the Weyl tensor are rather weak.

\begin{prop}
\label{prop_integrab}
In a five-dimensional Einstein spacetime admitting a  multiple WAND~$\lb$ with $\rhob \neq 0$, there exists an {\em optical structure} if any of the following conditions holds:
\begin{itemize}
 \item the spacetime is of type N, III or D
 \item the spacetime is of (genuine) type II, and $\rhob$ is either of the form (i) or (ii) of Theorem~\ref{th_GS}, or of the form (iii) {\em with $a=0$}.
\end{itemize}
In the case of type D spacetimes there exist in fact (at least) two optical structures if both geodesic multiple WANDs have a non-vanishing optical matrix. 
\end{prop}
In other words, all 5d algebraically special Einstein spacetimes with $\rhob\neq0$ admit an optical structure except, possibly, for those genuine type II spacetimes whose (unique) multiple WAND has an optical matrix of the form (iii) of Theorem~\ref{th_GS} with $a\neq 0$ (and is thus twisting). No examples of the latter subcase are presently known. Note, in particular, that the presence of an {\em expanding non-twisting} multiple WAND ensures the existence of an optical structure for all algebraically special Weyl types. In addition to the special type II twisting subcase (iii), also Kundt spacetimes ($\rhob=0$) evade the proposition. We observe finally that {\em the converse of Proposition~\ref{prop_integrab} does not hold}. This can be seen, e.g., by taking a black ring as a counterexample, since this is (in some region) a spacetime of type I$_i$ \cite{PraPra05} and yet admits an optical structure \cite{Taghavi-Chabert11}. 

\begin{proof}

That conditions (\ref{Dorth_integr_k_rho}) are met already follows from the previous discussion. Namely, a multiple WAND can always be chosen to be geodesic ($\kappa_{i}=0$) \cite{DurRea09}. Then, Theorem~\ref{th_GS} shows that the conditions on $\rho_{ij}$ are also satisfied for all types N and III, and for type II spacetimes with $\lb$ such that $\rhob$ is of the form (i) or (ii) of Theorem~\ref{th_GS}. Type II spaces whose multiple WAND has a $\rhob$ of the form (iii), instead, do not satisfy $\rho_{33}=\rho_{22}$ (unless $a=0$). 
This applies also to type D(iii), however in that case all such spacetimes are known (see Section~\ref{subsubsec_ex_iii}) and they all admit additional multiple WANDs $\tilde\lb$ and $\tilde{\tilde\lb}$ such that the corresponding $\tilde\rhob$ and $\tilde{\tilde\rhob}$ are both of the form (iii), but now with $a=0$ (so that (\ref{Dorth_integr_k_rho}) is satisfied by both ${\tilde\rhob}$ and $\tilde{\tilde\rhob}$ -- up to relabeling the spatial frame vectors, i.e. $2\leftrightarrow 4$).
From now on we shall thus exclude twisting genuine type II (iii) from the discussion.

Next, the canonical forms (i)--(iii) of $\rhob$ are compatible with using a parallelly transported frame \cite{OrtPraPra10}, so that one can always set $\M{i}{j}{0}=0$ \cite{OrtPraPra07}, so that both (\ref{Dorth_integr_k_rho}) and the first of (\ref{Dorth_integr_M}) are now satisfied. 

Finally, inserting the permitted forms of $\rho_{ij}$ into the Ricci identity (11k,\cite{OrtPraPra07}) (or, equivalently, the NP equation (A4,\cite{Durkeeetal10})) one finds that also the last two of (\ref{Dorth_integr_M}) are satisfied (cf. section IV.D.2 of \cite{OrtPraPra10} for an explicit proof in a special subcase; the general proof works similarly - in fact even with no need to specify the $r$-dependence of the Ricci rotation coefficients). One should observe that this argument does not work for Robinson-Trautman spacetimes, i.e. for a geodesic multiple WAND corresponding to (i) of Theorem~\ref{th_GS} with $a=0$ (i.e., $\rho_{ij}=b\delta_{ij}$), since (11k,\cite{OrtPraPra07}) is satisfied identically in that case. However, in 5d the Robinson-Trautman class reduces to Schwarzschild-like black holes \cite{PodOrt06} and one can construct explicitly an optical structure there.\footnote{Namely, the general metric can be written as $ds^2=-f(r)dt^2+f(r)^{-1}dr^2+r^2\Omega^2(dx^2+dy^2+dz^2)$, with $f(r)=k-\mu r^{-2}-\lambda r^2$ and 
$\Omega=\left[1+\frac{k}{4}(x^2+y^2+z^2)\right]^{-1}$. Then, taking the multiple WAND $\lb=f(r)^{-1}\pa_t+\pa_r$ and the spacelike frame vectors $\mb{2}=(r\Omega)^{-1}\pa_x$, $\mb{3}=(r\Omega)^{-1}\pa_y$, $\mb{4}=(r\Omega)^{-1}\pa_z$, one finds that ${\cal D}_{23}=\mbox{Span}\{\mb{2}+i\mb{3},\bl\}$ and ${\cal D}_{23}^\bot$, ${\cal D}_{24}=\mbox{Span}\{\mb{2}+i\mb{4},\bl\}$ and ${\cal D}_{24}^\bot$, and ${\cal D}_{34}=\mbox{Span}\{\mb{3}+i\mb{4},\bl\}$ and ${\cal D}_{34}^\bot$ are all integrable, so that there exist in fact three optical structures (of real index~1) associated with $\lb$ (plus their three complex conjugate ones). Another three (plus three) optical structures are obviously associated also with the second multiple WAND (obtained from $\lb$ just by time reflection).
\label{foot_RT}}

{It already follows from the remarks above} that for type D spacetimes with a multiple WAND falling into case (iii) of Theorem~\ref{th_GS} there exist two optical structures. The same conclusion clearly applies also to type D spacetimes having no multiple WAND of type (iii): indeed, the argument given above for type II (i)/(ii) spacetimes can be applied to each multiple WAND separately (provided both have an associated non-zero optical matrix). This concludes the proof.

\end{proof}

Eq.~(11k,\cite{OrtPraPra07}) is trivial also when $\rhob=0$, which explains why the proposition cannot be extended to the Kundt case; however, at least all ${\cal D}_{23}$,  
${\cal D}_{24}$ and ${\cal D}_{34}$ (see footnote \ref{foot_RT} for definitions) are integrable in that case.

\subsection{Comparison with the results of \cite{Taghavi-Chabert11}}

\label{subsec_TC}

\begin{table}[t]
  \begin{center}
  \begin{tabular}{|c|c|}
    \hline Notation of \cite{Taghavi-Chabert11} & Notation of this paper \\ \hline
    $\kappa$ & $\frac{1}{\sqrt{2}}(\kappa_2-i\kappa_3)$  \\
    $\mdigamma$ & $\kappa_4$  \\
    $\rho$ & $\frac{1}{2}[\rho_{22}+\rho_{33}+i(\rho_{23}-\rho_{32})]$ \\ 
    $\sigma$ & $\frac{1}{2}[\rho_{22}-\rho_{33}-i(\rho_{23}+\rho_{32})]$ \\
    $\psi$ & $\frac{1}{\sqrt{2}}(\rho_{42}-i\rho_{43})$  \\
    $\eta$ & $\frac{1}{\sqrt{2}}(\rho_{24}-i\rho_{34})$ \\
    $\yogh$ & $\rho_{44}$ \\ 
    $\chi$ & $\frac{1}{\sqrt{2}}(\M{2}{4}{0}-i\M{3}{4}{0})$ \\ 
    $\phi$ & $\frac{1}{\sqrt{2}}[\M{2}{4}{2}-\M{3}{4}{3}-i(\M{2}{4}{3}+\M{3}{4}{2})]$ \\ \hline
    $\Psi^0_{13}$ & $\frac{1}{2}[\Phi_{22}-\Phi_{33}-i(\Phi_{23}+\Phi_{32})]$ \\
    $\Psi^1_{13}$ & $\frac{1}{\sqrt{2}}(\Phi_{42}-i\Phi_{43})$ \\
    $\Psi^2_{13}$ & $\frac{1}{\sqrt{2}}(\Phi_{24}+i\Phi_{34})$ \\
    $\Psi^0_2$ & $\Phi_{44}$ \\
    $\Psi_2$ & $\frac{1}{2}[\Phi_{22}+\Phi_{33}+i(\Phi_{23}-\Phi_{32})]$ \\
    $\Psi_{14}$ & $\frac{1}{2}[\Psi_{224}'-\Psi_{334}'+i(\Psi_{324}'+\Psi_{234}')]$ \\
\hline
  \end{tabular}
    \caption{The Ricci rotation coefficients and Weyl frame components relevant to the discussion of the present paper are compared with the notation used in \cite{Taghavi-Chabert11}, provided one identifies the frame vectors $\mathbf m$ and $\mathbf u$ of \cite{Taghavi-Chabert11} with our $\frac{1}{\sqrt{2}}(\mb{2}+i\mb{3})$ and $\mb{4}$, respectively. In particular, the definition of ``algebraically special'' used in \cite{Taghavi-Chabert11} is stronger than ours in that it additionally assumes $\Psi^0_{13}=\Psi^1_{13}=\Psi^2_{13}=\Psi_{14}=0$ (plus a ``genericity'' condition).\label{tab_notation}}
  \end{center}
\end{table}

First, let us remark that the results of \cite{Taghavi-Chabert11} apply to a class of metrics larger than Einstein spacetimes: instead of requiring $R_{ab}=\frac{R}{5}g_{ab}$, a weaker condition (on the Cotton-York tensor) was imposed. However, we shall consider the results of \cite{Taghavi-Chabert11} as restricted to Einstein spaces, since only this is relevant to our paper. Note that the (complex) notation of \cite{Taghavi-Chabert11} is translated into ours in Table~\ref{tab_notation} (we give there only the quantities necessary for the present discussion, i.e. a subset of the Ricci rotation coefficients, all Weyl components of boost weight (b.w.) 0, and one component of b.w. -1).

Next, note that the class of ``algebraically special'' spacetimes in the sense of \cite{Taghavi-Chabert11} is narrower than ours: in addition to the vanishing of all positive b.w. Weyl components, the vanishing of the components (cf. Table~\ref{tab_notation}) $\Psi^0_{13}$, $\Psi^1_{13}$, $\Psi^2_{13}$, (of b.w. 0) and $\Psi_{14}$ (of b.w. -1) is also required.

Now, in our terminology and notation, Theorem~3.4 of \cite{Taghavi-Chabert11} can be rephrased as
\begin{itemize}
\item In a 5d Einstein spacetime whose associated Weyl tensor is algebraically special and additionally satisfies (in a null frame adapted to the multiple WAND $\lb$)
\be
 \Phi_{22}=\Phi_{33} , \quad \Phi_{23}=-\Phi_{32} , \quad \Phi_{42}=\Phi_{43}=\Phi_{24}=\Phi_{34}=0 , \qquad   \Psi_{224}'=\Psi_{334}' , \quad \Psi_{324}'=-\Psi_{234}' ,
 \label{TC_II}
\ee
and is otherwise ``generic'' (except that one can possibly have {\em either} $\Phi_{44}=0$ {\em or} $\Phi_{22}=\Phi_{33}=\Phi_{23}=\Phi_{32}=0$, see subsection~3.4.2 of \cite{Taghavi-Chabert11} and Table~\ref{tab_notation}) the distributions (\ref{D}), (\ref{D_orth}) define an optical structure.
\end{itemize}

This result clearly applies only to a subset of type II Einstein spacetimes (as noted also in \cite{Taghavi-Chabert11}). In particular, an assumption on negative b.w. component is also necessary. It was derived by a detailed study of the Bianchi identities.

For the special case of type D spacetimes, Corollary~3.10 of \cite{Taghavi-Chabert11} follow immediately, this can be rephrased as:
\begin{itemize}
\item In a 5d Einstein spacetime whose associated Weyl tensor is of type D and additionally satisfies (in a null frame adapted to two multiple WANDs $\lb$ and $\nb$)
\be
 \Phi_{22}=\Phi_{33} , \quad \Phi_{23}=-\Phi_{32} , \quad \Phi_{42}=\Phi_{43}=\Phi_{24}=\Phi_{34}=0 , 
 \label{TC_D}
\ee
and is otherwise ``generic'' ($\Phi_{22}\neq 0$, $\Phi_{23}\neq 0$, $\Phi_{44}\neq 0$) the distributions (\ref{D}), (\ref{D_orth}) and the corresponding distributions with $\lb$ replaced by $\nb$ define two optical structures.
\end{itemize}

The above result applies, for example, to the Myers-Perry spacetime (see \cite{Taghavi-Chabert11} for more comments), cf. the Weyl tensor components given in \cite{PraPraOrt07}. However, in this form it does not apply, e.g., to the 5d Schwarzschild spacetime, since one has $\Phi_{23}=0$ (and $\Phi_{44}=\Phi_{22}$) in that case, so the ``genericity'' assumption is clearly violated. On the other hand, 
by using also the Ricci identity, we have seen above that in all type D spacetimes there exist two optical structures, provided both multiple WANDs have a non-zero optical matrix. The extra assumptions (\ref{TC_D}) and the ``genericity'' condition thus appear to be unnecessary (at least in the non-Kundt case).

Similarly, the comments in section~3.4.2 of \cite{Taghavi-Chabert11} demonstrate that the proof of \cite{Taghavi-Chabert11} cannot be extended to type III/N spacetimes in general. However, by studying also consequences of the Ricci identity we have seen above that the conclusion remains true.

\section{Examples}

\label{sec_examples}

{In this section we present explicit examples of type II (or D) Einstein spacetimes for each of the possible cases (i), (ii) and (iii) of Theorem~\ref{th_GS} and a counterexample to  the converse of Theorem~\ref{th_GS}. In addition to some familiar solutions, we shall discuss some examples which have not been considered in the context of algebraically special solutions, notably Kaluza-Klein bubbles arising from analytic continuation of the Myers-Perry solution.  

\label{Sec_examples}

\subsection{Optical matrix of the form \eqref{5dform1}}

\label{sec_form_i}

As already mentioned, examples of Einstein spacetimes belonging to the case (i) of Theorem~\ref{th_GS} are  Myers-Perry black hole and, more generally,
all non-degenerate (i.e. $\det\rho\not= 0$)  Einstein Kerr-Schild metrics with Minkowski or (A)dS background \cite{OrtPraPra09,MalPra11GKS}.

Another example belonging to this class is the 5d Kaluza-Klein bubble obtained by analytic continuation of a Myers-Perry solution \cite{Dowkeretal95}. (The original Kaluza-Klein bubble of Ref. \cite{Witten82} will be discussed below.) We start with the singly spinning Myers-Perry metric in the form
\be
{\rm d} s^2=  \frac{r^2\rho^2}{\Delta} {\rm d}r^2-{\rm d}t^2+\rho^2 {\rm d} \theta^2+ (r^2+A^2) \sin^2 \theta {\rm d} \phi^2 + r^2
\cos^2\theta {\rm d} \psi^2   
 + \frac{{r_0}^2}{\rho^2} ({\rm d}t+A \sin^2 \theta {\rm d}\phi 
 )^2,
\ee
where
\BEA
 \rho^2&=&r^2+A^2\cos^2\theta, \qquad \Delta=r^2(r^2+A^2)-r_0^2r^2 .
\EEA
After performing a transformation  $t \rightarrow i \chi, \theta  - \frac{\pi}{2} \rightarrow i \tau, \ \psi \rightarrow i \sigma,\ A \rightarrow i \alpha $  we  arrive to a metric
\be
{\rm d} s^2=  \frac{r^2\rho^2}{\Delta} {\rm d}r^2- \rho^2 {\rm d} \tau^2+ {\rm d}\chi^2+(r^2-\alpha^2) \cosh^2 \tau {\rm d} \phi^2 +   r^2\sinh^2\tau {\rm d} \sigma^2 
 - \frac{{r_0}^2}{\rho^2} ({\rm d}\chi+\alpha \cosh^2 \tau {\rm d}\phi )^2,
\ee
with
\BEA
 \rho^2&=&r^2+\alpha^2\sinh^2\tau 
 , \qquad \Delta=r^2(r^2-\alpha^2) 
 -r_0^2r^2 .
\EEA

By studying the higher-dimensional Bel-Debever criteria \cite{Ortaggio09} one finds that this metric is of type D with
the multiple geodesic WANDs of the form
\be
\ell_{\pm a} {\rm d} x^a= \pm  \frac{r^2+\alpha^2 \sinh^2 \tau}{(r^2-\alpha^2) \cosh \tau } {\rm d} \tau + \frac{\alpha}{r^2-\alpha^2} {\rm d} \chi - {\rm d} \phi .
\ee
We choose $\bl$ and $\bn$ coinciding with $\bl_{+}$ and $\bl_{-}$, respectively (up to a possible rescaling), and the rest of the frame as
\bea
 m_{(2)a}d x^a=\gamma^{-1} {\rm d} r, \qquad m_{(3)a}d x^a= \gamma  {\rm d} \chi + \alpha \gamma \cosh^2 \tau {\rm d} \phi , \qquad 
m_{(4)a}d x^a= r \sinh \tau  {\rm d} \sigma ,
\eea
with $\gamma=\sqrt{\frac{r^2-{r_0}^2-\alpha^2}{\nu}}$, where $\nu=r^2+\alpha^2 \sinh^2 \tau$.  Then the optical matrix is 
\be
 \rho_{ij} =  -\frac{1}{(r^2-\alpha^2)}\left( \begin {array}{ccc}  \frac{\alpha^2 \sinh \tau}{\nu } & \frac{- \alpha r}{\nu }  & 0
\\\noalign{\medskip} \frac{ \alpha r}{\nu }  &  \frac{\alpha^2 \sinh \tau}{\nu }  & 0
\\\noalign{\medskip} 0 & 0 & \frac{1}{  \sinh \tau}
\end {array}
 \right).
\ \ \ \, \label{KKbubble_rho}
 \ee
 This corresponds to case (i) of Theorem~\ref{th_GS} with
 \be
  b = - \frac{\alpha^2 \sinh \tau}{\nu (r^2 -\alpha^2)}, \qquad a = - \frac{r}{\alpha \sinh \tau}.
 \ee
 
\subsection{Optical matrix of the form \eqref{5dform2}}

\label{subsubsec_ex_ii}

A large class of Ricci-flat solutions admitting a multiple WAND with optical matrix of the form \eqref{5dform2} {(case (ii) of Theorem~\ref{th_GS})}, {such as the Kerr black string,} is given by taking the product of a 4d Ricci-flat algebraically special solution with a flat 5th direction. {Similarly,} examples with non-vanishing cosmological constant can be obtained by taking the warped product of a 4d algebraically special Einstein spacetime with a 5th direction, i.e., a solution of the form
\be
 ds^2 = w(y)^2 ds_4^2 + dy^2 ,
\ee
for a suitable choice of the function $w(y)$. 
In this case, $ds^2$ will be of the same (special) type as $ds_4^2$ \cite{OrtPraPra11}, so that several examples of any algebraically special type can be easily constructed. See \cite{OrtPraPra10} for type N/III examples.

\subsection{Optical matrix of the form \eqref{5dform3}}

\label{subsubsec_ex_iii}

{We are not presently aware of any such examples for the genuine type II. By contrast, type D spacetimes with optical matrix of the form \eqref{5dform3} {(case (iii) of Theorem~\ref{th_GS})} coincide (as shown in Section~\ref{sec_D_iii}) with the algebraically special metrics admitting a non-geodesic multiple WANDs. These were all given in Theorem~3 of \cite{DurRea09} and consist of two subfamilies, which can thus be presented here (together with a suitable choice of a multiple WAND) as examples falling into case (iii) of Theorem~\ref{th_GS}.}

The first subfamily of the metrics of \cite{DurRea09} is given by the direct products $dS_3 \times S^2$ and $AdS_3 \times H^2$, which can be written in a unified form as
\beqn
 & & ds^2=\Omega^2(-dt^2+dx^2+dy^2)+\Sigma^2(dz^2+dw^2) , \\
 & & \mbox{with } \ \Omega^{-1}=1+\frac{\lambda}{2}(-t^2+x^2+y^2) , \quad \Sigma^{-1}=1+\lambda(z^2+w^2) ,
\eeqn  
where $\lambda\neq 0$ is (proportional to) the cosmological constant. 
Ref. \cite{GodRea09} observed that {\it any} null vector field tangent to $(A)dS_3$ is a multiple WAND, e.g.
the null vector
\be
 \ell_ad x^a=\left(1+\frac{\alpha^2}{4}\right)d t+\left(1-\frac{\alpha^2}{4}\right)d x+\alpha d y . 
 \label{l_S2_AdS3}
\ee
This is a geodesic, affinely parametrized multiple WAND for any $\alpha=\alpha(z,w)$. Using the frame 
\beqn
 & & n_ad x^a=\frac{1}{2}\Omega^2(-d t+d x) , \quad m_{(2)a}d x^a=\Omega\left[\frac{\alpha}{2}(d t-d x)+d y\right] , \nonumber \\
 & & m_{(3)a}d x^a=\Sigma d z, \quad m_{(4)a}d x^a=\Sigma dw ,
\eeqn
one finds that $\rhob$ has the only non-zero components
\be
 \rho_{22}=\frac{\lambda}{4\Omega}\left[-4(t+x)-4y\alpha+(-t+x)\alpha^2\right] , \quad \rho_{23}=\frac{\alpha_{,z}}{\Omega\Sigma} , \quad \rho_{24}=\frac{\alpha_{,w}}{\Omega\Sigma} .
\ee
For generic $\alpha(z,w)$, $\rho_{(ij)}$ has one vanishing eigenvalue and two non-vanishing and unequal eigenvalues. This implies that the canonical form must be \eqref{5dform3} with $a\ne 0$ {(indeed this matrix $\rhob$ can obtained from~(\ref{5dform3}) using spins)}.
For the special choice $\alpha=$const the form of $\rhob$ degenerates to \eqref{5dform3} with $a=0$ (i.e., with zero twist). In this case, it is then easy to see that ${\cal D}=\mbox{Span}\{\mb{3}+i\mb{4},\bl\}$ together with its orthogonal complement ${\cal D}^\bot$ defines an optical structure, {in agreement with Proposition~\ref{prop_integrab} (the same is true for the null vector obtained from $\lb$ by time-reflection, i.e., $t\to-t$, giving rise to another optical structure)}.

The second subfamily of \cite{DurRea09} is given by an analytical continuation of the  5d Schwarzschild solution
(generalized to include a cosmological constant $\Lambda$ and planar or hyperbolic symmetry), and can be written as
\beqn
 & & ds^2=f(r)dz^2+f(r)^{-1}dr^2+r^2\Omega^2(-dt^2+dx^2+dy^2) , \\
 & & \mbox{with } \ f(r)=k-\mu r^{-2}-\lambda r^2, \quad \Omega^{-1}=1+\frac{k}{4}(-t^2+x^2+y^2) ,
\eeqn  
where $\mu\neq0$, $k \in \{1,0,-1 \}$, $\lambda$ is (proportional to) the cosmological constant, and the coordinate $r$ takes values so that $f(r)>0$. 
A special case is the Kaluza-Klein bubble of { \cite{Witten82}}. In all of these cases, any geodesic null vector field tangent to the 3d space of constant curvature is a geodesic multiple WAND  and the optical matrix has canonical form \eqref{5dform3}. 
Let us choose 
a null vector
\be
 \ell_ad x^a=\left(1+\frac{\alpha^2}{4}\right)d t+\left(1-\frac{\alpha^2}{4}\right)d x+\alpha d y ,\label{l_Schw}
\ee
where $\alpha=\alpha(z,r)$, that is a geodesic, affinely parametrized multiple WAND. Using the frame 
\beqn
 & & n_ad x^a=\frac{1}{2}r^2\Omega^2(-d t+d x) , \quad m_{(2)a}d x^a=r\Omega\left[\frac{\alpha}{2}(d t-d x)+d y\right] , \nonumber \\
 & & m_{(3)a}d x^a=f(r)^{1/2}d z, \quad m_{(4)a}d x^a=f(r)^{-1/2}dr ,
\eeqn
one finds that $\rhob$ has the only non-zero components
\be
 \rho_{22}=\frac{k}{8\Omega r^2}\left[-4(t+x)-4y\alpha+(-t+x)\alpha^2\right] , \quad \rho_{23}=\frac{\alpha_{,z}}{r\Omega f(r)^{1/2}} , \quad \rho_{24}=\frac{\alpha_{,r} f(r)^{1/2}}{r\Omega} .
\ee

Similarly as above, in general $\rhob$ can be cast into the canonical form \eqref{5dform3} with $a\ne 0$. In the special case $\alpha=$const, i.e. a hypersurface-orthogonal geodesic multiple WAND,  the form of $\rhob$ reduces to \eqref{5dform3} with $a=0$ (additionally, also $b=0$ when $k=0$). With this choice, ${\cal D}=\mbox{Span}\{\mb{3}+i\mb{4},\bl\}$ together with its orthogonal complement ${\cal D}^\bot$ defines an optical structure (and similarly for the time-reflection of $\lb$).

\subsection{Counterexample to the converse of Theorem~\ref{th_GS}}

\label{subsec_counterex}

Finally we present a counterexample to  the converse of Theorem~\ref{th_GS}. We shall exhibit an Einstein spacetime with a null geodesic vector field whose optical matrix has the form (iii) of Theorem~\ref{th_GS} but is not a multiple WAND.

Consider a 4d cylindrical Newman-Tamburino solution (see Eq. (26.23) in \cite{Stephanibook}) \be
{\rm d}s^2 = r^2 {\rm d} x^2 + x^2 {\rm d} y^2 - \frac{4 r}{x } {\rm d} u {\rm d} x - 2 {\rm d} u {\rm d} r + x^{-2} \left(c + \ln (r^2 x^4)\right) {\rm d} u^2 , \label{NTmetric}
\ee
{where $c$ is a constant.} This is a type I Ricci-flat spacetime with  $\l_adx^a = {\rm d} u$ being a geodesic principal null direction with optical matrix of the form diag$(b,0)$.
A direct product  of this spacetime with a flat dimension is  a five-dimensional type I$_i$ Ricci flat\footnote{A non-Ricci-flat Einstein space can be obtained by taking a warped product with the fifth dimension. Note that under special circumstances a direct/warp product of an Einstein type I spacetime can lead to an Einstein type D spacetime, however, this occurs only in dimension $d\geq 6$, see \cite{OrtPraPra11} for details. } spacetime with ${\rm d} u$ 
being a WAND \cite{OrtPraPra11} but not a multiple WAND. The corresponding form of the optical matrix is diag$(b,0,0)$, i.e. case (iii) of Theorem~\ref{th_GS} with $a=0$. Therefore {\it the existence of a null geodesic congruence whose optical matrix takes the canonical form (iii) of Theorem 1 is not a sufficient condition for the spacetime to be algebraically special}.

\section{Discussion}

\label{sec_discussion}

{We have presented an extension to five dimensions (Theorem~\ref{th_GS}) of the ``shearfree part'' of the 4d Goldberg-Sachs theorem. Combined with the result of Ref. \cite{DurRea09} on the the ``geodesic part'' this provides a five-dimensional generalization of the Goldberg-Sachs theorem for Einstein spacetimes. 

We have given explicit examples corresponding to each case of Theorem~\ref{th_GS}, so this result is sharp. However, only {\em necessary} conditions for a spacetime to be algebraically special have been obtained. That our conditions cannot be sufficient has been demonstrated by an example (Section~\ref{subsec_counterex}) of a 5d Einstein spacetime which is not algebraically special yet has geodesic null vector field whose optical matrix takes the form (iii) of Theorem~\ref{th_GS}. Hence form (iii) is not sufficient for a geodesic null vector field to be a multiple WAND. An interesting open question is whether forms (i) and (ii) are sufficient. 

Our analysis has demonstrated some restrictions on the possible structure of algebraically special Weyl tensors of Einstein spacetimes, as expressed in terms of the permitted  ``spin types'' defined in Ref. \cite{Coleyetal12}. Our results will lead to a simplification of the Einstein and Bianchi equations for algebraically special solutions, and it is  hoped that this will be helpful in constructing and analyzing new explicit solutions of the 5d Einstein equations in vacuum.

As discussed in Section \ref{Sec_OC} any matrix satisfying the optical constraint can be brought to one of the forms (\ref{5dform1}), (\ref{5dform2}) or, with $a=0$, (\ref{5dform3}) by an appropriate choice of orthonormal basis. Conversely, each of these forms satisfies the optical constraint. The form (\ref{5dform3}) with $a \ne 0$ violates the optical constraint. But, as discussed, in the examples known to us (Sections~\ref{sec_form_i}--\ref{subsubsec_ex_iii}), solutions with an optical matrix of this type always admit a continuous family of multiple WANDs (and are thus of type D),
and this family contains a {geodesic} multiple WAND with optical matrix of the form (\ref{5dform3}) with $a=0$, which obeys the optical constraint. {In fact,} if one restricts to type D spacetimes, then the solutions admitting a geodesic multiple WAND which violates the optical constraint 
are all explicitly known (cf. Section~\ref{sec_D_iii}) and consist of the examples of Sections~\ref{sec_form_i}--\ref{subsubsec_ex_iii}. 
A question thus remains as to whether there exist 5d Einstein spacetimes of genuine type II whose unique multiple WAND (is geodesic, twisting and) violates the optical constraint (i.e., falls into class (iii) with $a\neq 0$). If the answer were negative then Theorem~\ref{th_GS} could be reformulated as the statement that a 5d algebraically special Einstein spacetime that is not conformally flat must admit a geodesic multiple WAND that satisfies the optical constraint.

We have compared our results with an alternative 5d generalization of the Goldberg-Sachs theorem given in Ref.\cite{Taghavi-Chabert11}, which is based on a stronger definition of ``algebraically special''. (That reference also obtained only necessary conditions.) Some of the results of Ref. \cite{Taghavi-Chabert11} have been  strengthened using the Ricci identity. In particular, as a consequence of Theorem~\ref{th_GS} we have shown the existence of an optical structure for a large class of algebraically special solutions: only Kundt spacetimes  and non-Kundt genuine type II spacetimes falling in case (iii) of Theorem~\ref{th_GS} with $a,b\neq 0$ (of which no example is known to the authors) possibly evade such a conclusion. Finally, our work also contains a few results that hold in any higher dimensions. An extended discussion in $d$ dimensions will be presented elsewhere.

\hspace{1cm}

\noindent
{\bf Acknowledgments}\\ { MO is grateful to Lode Wylleman for useful discussions.}
AP and VP would like to thank DAMTP, University of Cambridge, for its hospitality while part of
this work was carried out. {MO,} AP and VP also acknowledge support from research plan { RVO: 67985840} 
and research grant no P203/10/0749. HSR is supported by a Royal Society University Research Fellowship and European Research Council grant ERC-2011-StG 279363-HiDGR. 

\appendix
\section{Hypersurface orthogonal multiple WAND $\bl$ with $\Phia_{ij} \neq 0$}
\label{app}

In some cases, constraints on the optical matrix can be obtained in arbitrary dimension. 
Here, let us  discuss a case with a hypersurface orthogonal multiple WAND $\bl$ with $\Phia_{ij} \neq 0$.
{Because hypersurface orthogonal and null,} $\bl$ is automatically geodesic {and twistfree}. Thus we have 
\be
 \kappa_i = A_{ij}=0 ,
\ee
{so that $\rho_{ij}=S_{ij}$.}
Eq.~\eqref{B8asym} reduces to
\be
\label{eqn:hyporthog}
 \rho \Phi^A_{ij} - \Phi_{ik} \rho_{kj} + \Phi_{jk} \rho_{ki} = 0 .
\ee

Acting with $\tho$ and using the Bianchi identity \eqref{A2} and Sachs equation (\ref{Sachs}) gives
\be
 -\left( \rho^2/2 + \rho_{kl} \rho_{kl} \right) \Phi^A_{ij} - 2\rho \Phi^A_{[i|k} \rho_{k|j]} + 8 \Phi^A_{[i|k} (\rho^2)_{k|j]} + 4 \Phi^S_{[i|k} (\rho^2)_{k|j]} = 0,
\ee
where $(\rho^2)_{ij} = \rho_{ik} \rho_{kj}$.  
Note that this equation can be also obtained without differentiation using \eqref{B4}, \eqref{eqn:hyporthog}
 and an identity from table \ref{tab:weyl}. 
The final term is the  commutator $2[\Phib^S,\rhob^2]$, which can be written in terms of $[\Phib^S,\rhob]$ and thus eliminated using equation (\ref{eqn:hyporthog}), giving
\be
  -\left( \rho^2/2 + \rho_{kl} \rho_{kl} \right) \Phi^A_{ij} + 2\rho \Phi^A_{[i|k} \rho_{k|j]} + 4 \Phi^A_{[i|k} (\rho^2)_{k|j]} - 4 \rho_{ik} \Phi^A_{kl} \rho_{lj} = 0.
 \ee
{From now on} {we use a basis in which $\rho_{ij}$ is diagonal, i.e. $\rho_{ij}=\diag (\rho_{2},\rho_{3},\dots)$}. The above equation thus reduces to 
\be
 \left( -\rho^2/2- \rho_{kl} \rho_{kl} + \rho(\rho_{(i)} + \rho_{(j)}) + 2(\rho_{(i)}-\rho_{(j)})^2 \right) \Phi^A_{(i)(j)}=0,
\ee
where parentheses indicate no summation over repeated indices $i,j$.

Now let us assume that $\Phi^A_{ij} \ne 0$ for some $i$ and $j$. Without loss of generality, we may assume that $\Phi^A_{23} \ne 0$. Then the above equation implies that
\be
 -\rho^2/2- \rho_{kl} \rho_{kl} + \rho(\rho_2 + \rho_3) + 2(\rho_2-\rho_3)^2 =0.
\ee
This can be rewritten as
\be
\label{eqn:evaleq}
 ({\rm tr} (\hat{\rho}))^2 + 2 {\rm tr} (\hat{\rho}^2) = 3 (\rho_2 - \rho_3)^2,
\ee
where $\hat{\rho}=\diag (\rho_4, \rho_5 \ldots)$. Although we have derived this equation in a particular basis, it involves only invariant quantities (eigenvalues of $\rho$) hence it must be basis-independent. 

The consequences of this equations can be easily seen by introducing a parallelly transported frame (in which $\tho=D=\pa_r$)\footnote{Note that choosing a parallelly transported frame is compatible with {using an} eigenframe of $\rhob$ \cite{PraPra08,OrtPraPra10}.\label{compatible}} and looking at the $r$-dependence of the various terms. First, let us recall that in the non-twisting case we have \cite{PraPra08} (see also \cite{OrtPraPra10} and Appendix~D of \cite{OrtPraPra09})
\be
 \rho_i=\frac{s^0_i}{1+rs^0_i} ,
 \label{rho_i_nontwist}
\ee
which can also be rewritten as $\rho_i=1/(r-b_i)$ when $s^0_i\neq 0$. There are several cases to consider. First assume $\rho_2=\rho_3$. Then, since the LHS of (\ref{eqn:evaleq}) is positive definite, we obtain $\hat{\rho}_{ij}=0$, i.e., $\rho_4=\rho_5 = \ldots = 0$.

Now assume $\rho_2$ and $\rho_3$ are distinct and non-zero. Then, plugging~(\ref{rho_i_nontwist}) into (\ref{eqn:evaleq}) and looking at the structure of poles  of the RHS and LHS we first find that
{ any non-vanishing $\rho_\alpha$ ($\alpha\ge 4$) must equal} $\rho_2$ or $\rho_3$, but then we arrive to a contradiction. 

The final case to consider is (without loss of generality) $\rho_2 \ne 0$, $\rho_3=0$. In this case, looking at the structure of poles of the RHS and of the LHS reveals that (permuting the basis if necessary) $\rho_4=\rho_2$, and $\rho_5 = \rho_6 = \ldots = 0$. This satisfies (\ref{eqn:evaleq}) without further restriction.

We see that in both allowed cases, $\rho_{ij}$ has two equal eigenvalues and $d-4$ vanishing eigenvalues. In conclusion:
\begin{prop}
 \label{prop_PhiA}
For non-twisting type II Einstein spacetimes with $\Phi^A_{ij} \ne 0$,  the eigenvalues of $\rhob$ are $\rho/2,\rho/2,0,\ldots, 0$.
\end{prop}

Note, in particular, that $\rho_{ij}$ is degenerate, and shearing unless identically zero.\footnote{This comment applies for $d>4$. Remember however that for non-twisting, expanding solutions in four dimensions one has $\Phi^A_{ij}=0$ (which is equivalent to $\Psi_2$ being real in the four-dimensional Newman-Penrose notation). This can be seen using the NP formalism to fix the $r$-dependence of the Weyl components, Ricci coefficients and derivative operators, and then using (7.32f) of \cite{Stephanibook}.  In fact these solutions reduce to the Robinson-Trautman class thanks to the 4d Goldberg-Sachs theorem.}

\end{document}